%% file: main.tex
\documentclass[12pt]{article}

\usepackage{amssymb,amsmath,amsfonts, bbm,eurosym,geometry,ulem,graphicx,color,setspace,sectsty,comment,footmisc,pdflscape,subfig,array}
\usepackage{multirow, makecell, booktabs}
\usepackage{csquotes}
\usepackage{graphicx, tikz}
\usetikzlibrary{shapes.geometric,fit,shapes, positioning, calc}
\usepackage{natbib}
\usepackage[title]{appendix}
\usepackage{hyperref}
\hypersetup{colorlinks,linkcolor={blue},citecolor={blue}, urlcolor={black}}  
\usepackage{xurl}
\usepackage{pdflscape}
\usepackage{caption}[bf]
\newcolumntype{L}[1]{>{\raggedright\let\newline\\arraybackslash\hspace{0pt}}m{#1}}
\newcolumntype{C}[1]{>{\centering\let\newline\\arraybackslash\hspace{0pt}}m{#1}}
\newcolumntype{R}[1]{>{\raggedleft\let\newline\\arraybackslash\hspace{0pt}}m{#1}}

\usepackage{etoolbox}
\makeatletter

\pretocmd{\NAT@citex}{%
  \let\NAT@hyper@\NAT@hyper@citex
  \def\NAT@postnote{#2}%
  \setcounter{NAT@total@cites}{0}%
  \setcounter{NAT@count@cites}{0}%
  \forcsvlist{\stepcounter{NAT@total@cites}\@gobble}{#3}}{}{}
\newcounter{NAT@total@cites}
\newcounter{NAT@count@cites}
\def\NAT@postnote{}

\def\NAT@hyper@citex#1{%
  \stepcounter{NAT@count@cites}%
  \hyper@natlinkstart{\@citeb\@extra@b@citeb}#1%
  \ifnumequal{\value{NAT@count@cites}}{\value{NAT@total@cites}}
    {\ifNAT@swa\else\if*\NAT@postnote*\else%
     \NAT@cmt\NAT@postnote\global\def\NAT@postnote{}\fi\fi}{}%
  \ifNAT@swa\else\if\relax\NAT@date\relax
  \else\NAT@@close\global\let\NAT@nm\@empty\fi\fi
  \hyper@natlinkend}
\renewcommand\hyper@natlinkbreak[2]{#1}

\patchcmd{\NAT@citex}
  {\ifNAT@swa\else\if*#2*\else\NAT@cmt#2\fi
   \if\relax\NAT@date\relax\else\NAT@@close\fi\fi}{}{}{}
\patchcmd{\NAT@citex}
  {\if\relax\NAT@date\relax\NAT@def@citea\else\NAT@def@citea@close\fi}
  {\if\relax\NAT@date\relax\NAT@def@citea\else\NAT@def@citea@space\fi}{}{}

\makeatother

\newcommand{\footremember}[2]{%
    \footnote{#2}
    \newcounter{#1}
    \setcounter{#1}{\value{footnote}}%
}

\newcommand{\footrecall}[1]{%
    \footnotemark[\value{#1}]%
} 

\usepackage[textsize=tiny,textwidth=1.4cm]{todonotes}

\begin{document}
\begin{titlepage}
\title{Understanding the Excess Bond Premium}
\author{ Kevin Benson\footremember{rbc}{Royal Bank of Canada Capital Markets. New York, NY and Toronto, ON, Canada.} \ \
Ing-Haw Cheng\footremember{uoft}{Rotman School of Management, FinHub Financial Innovation Lab, University of Toronto. 105 St. George Street, Toronto, ON, M5S 3E6, Canada.}\thanks{Corresponding author. The authors thank Shivesh Prakash for excellent research assistance. The authors also thank the Royal Bank of Canada for generous funding of the FinHub Financial Innovation Lab. Ing-Haw Cheng thanks the Social Science and Humanities Research Council of Canada (SSHRC) for generous funding through SSHRC Insight Grant \#435-2024-0365. The views expressed in this article are those of the authors and do not necessarily reflect those of the Royal Bank of Canada.} \ \
 John Hull\footrecall{uoft} \ \
 Charles Martineau\footrecall{uoft} \\
 Yoshio Nozawa\footrecall{uoft} \ \
 Vasily Strela\footrecall{rbc} \ \ 
  Yuntao Wu\footrecall{uoft} \ \
 Jun Yuan\footrecall{rbc} \footrecall{uoft}
 }
\date{%
  \today
  \\
  \vspace{.5cm}
  PRELIMINARY
}
\maketitle

\setstretch{1}
\begin{abstract}
\noindent We study the drivers of the \citet{GZ:12} excess bond premium (EBP) through the lens of the news. The monthly attention the news pays to 180 topics \citep{bkmx:21} captures up to 80\% of the variation in the EBP, and this component of variation forecasts macroeconomic movements. Greater news attention to financial intermediaries and crises tends to drive up the EBP and portend macroeconomic downturns, while greater news attention to politics and science tends to drive down the EBP. Attention-based estimates of EBP largely drive out the forecast power of direct sentiment measures for macroeconomic fluctuations and predict the business cycle going back to the early 1900's. Overall, we attribute predictive variation about the EBP for macroeconomic movements to variation in news attention to financial intermediaries, crises, and politics.

\vspace{.25in}
\vspace{0in}

\bigskip
\end{abstract}
\setcounter{page}{0}
\thispagestyle{empty}
\end{titlepage}
\setstretch{1.5}

\noindent What drives the excess bond premium (EBP), and why does it predict macroeconomic fluctuations? The EBP equals the difference between corporate credit spreads and spreads implied from measured default risk averaged across issuers. Since the work of \citet{GZ:12}, economists have increasingly recognized that elevated levels of these differences tend to lead macroeconomic downturns as they may reflect deterioration in the health of the financial sector. However, the drivers of the EBP have been difficult to pin down because the EBP reflects, by its nature, a ``non-fundamental''  component of credit spreads. The question of what drives variation in the EBP and its forecast power for business cycle fluctuations thus remains open.

This paper addresses this question by assigning variation in EBP to attention to different topics in the news with the following goals in mind. First, we can quantify how much variation in EBP is associated with economically interpretable news categories and whether results differ for the ``fundamental'' default-risk component of credit spreads. Moreover, we can ask whether the movements attributable to those categories flow through to forecast macroeconomic movements. Second, we can study whether that forecast power remains in periods farther back in history and removed from the 2007-2009 Great Financial Crisis, extending the out-of-sample test of the forecast power of EBP and its news-driven components. In particular, we use methods from \citet{bkmx:21} and \citet{nvix} and a comprehensive dataset of Wall Street Journal articles to estimate EBP from attention to different topics in the news. We decompose this estimate into topic-level components and extend our sample back to the early 1900's.

We have three main findings. First, greater news attention to financial intermediaries drives up the EBP and portends bad news for the macroeconomy, consistent with theories and existing explanations for the EBP \citep{GZ:12}. Second, there is significant variation not directly related to financial intermediaries that contain forecast power for macroeconomic fluctuations, particularly in politics and plausibly sentiment-related topics. Indeed, our estimate of EBP often drives out the forecast power of direct measures of news sentiment. Several patterns also appear distinct from those driving the default-risk component of credit spreads. Finally, we confirm the forecast power of our attention-driven EBP for macroeconomic fluctuations going back to the early part of the 20th century.

Section \ref{sec:methodology} describes and quantitatively validates our methodology to estimate the excess bond premium using text data. We estimate $\hat{EBP}$, the component of EBP explained by variation in news attention, using \citet{bkmx:21}'s topic model of 180 distinct news topics. We show that $\hat{EBP}$ captures up to 86\% of the variance in EBP during our 15-year training period spanning 1973-1987. From 1988 onwards, the $\hat{EBP}$ captures 78\% of EBP's variance when re-estimating the topic weights contemporaneously with new data using an expanding window approach. A more conservative approach that re-estimates the model with a 1-month delay, and thus constructs $\hat{EBP}$ each month using stale model weights, generates a news-driven EBP that still captures 45\% of the variation in EBP.

Section \ref{sec:understandingebp} breaks down the variation and forecast power of $\hat{EBP}$ into economically interpretable news topics. We first show that $\hat{EBP}$ contains significant forecast power for unemployment, industrial production, and recession risks.  Quantitatively, the effect of a one-point move in $\hat{EBP}$ has a similar magnitude on forecasts of macroeconomic movements as a one-point movement in EBP itself. Moreover, analysis of \citet{SHAP} SHAP (SHaply Additive exPlanations) values shows that the contribution of $\hat{EBP}$ to the forecast model is greater than that of the residual difference of $EBP$ and $\hat{EBP}$, although this residual difference retains some forecasting power in our estimates.

Our first main finding is that greater news attention to financial intermediaries and crisis-related topics (both financial and non-financial) tend to drive up $\hat{EBP}$, and that this variation translates into forecast power for the business cycle. These news topics are relatively more important for $\hat{EBP}$ than for the default-risk component of credit spreads, where they typically have a smaller variance share or attention weight.

Our second main finding is that $\hat{EBP}$ contains significant topic-level variation not directly related to financial intermediaries in ways that differ from the default-risk component of credit spreads yet also translate into forecast power. News about politics and industry are top topics that tend to drive down $\hat{EBP}$, even though political news is roughly neutral for the default-risk component. News topics associated with negative sentiment receive a greater weight for $\hat{EBP}$ than in the default-risk component. Finally, general news about financial markets and the economy drives down $\hat{EBP}$, whereas it drives up and is the most important contributor of variance to the default-risk component.

Non-financial-intermediary topic variation in $\hat{EBP}$ flows through to forecast power for macroeconomic movements for several, but not all, of the topics. News about crises, politics, and negative sentiment drive variation in $\hat{EBP}$, and that same predicted variation contributes to forecast power for macroeconomic fluctuations. However, variation in $\hat{EBP}$ associated with news about industry and a residual category of ``Other'' news does not flow through to forecast power, even though news in those topics, particularly ``Other,'' drive a significant amount of variation in $\hat{EBP}$. 

To explore the role of sentiment, we construct three empirical measures of sentiment to test whether they add additional forecast power. We measure raw market sentiment, the weighted average of topic-level sentiment, and the fitted value of EBP to raw market sentiment. We use the \citet{LM:11} dictionary of sentiment-charged words to construct our measures. We find that each of these measures forecasts our same macroeconomic indicators, but that the inclusion of $\hat{EBP}$ drives out the forecast power of these measures when including both sets of variables as predictors. These results suggest that sentiment-related forecast power flows through our attention-based measures rather than measures based directly on sentiment-charged words.

Section \ref{sec:historical} explores our third main finding, that $\hat{EBP}$ predicts macroeconomic fluctuations in the longer time series. We train a model to predict EBP using news topics with data from 1973 to the present, and then project $\hat{EBP}$ back to the start of the 20th century in the spirit of \citet{nvix}. We show that $\hat{EBP}$ predicts a host of macroeconomic indicators in the historical non-training period spanning the start of each series in the early 20th century to 1972, validating the predictive power of $\hat{EBP}$ for the macroeconomy.

Section \ref{sec:robust} explores robustness. Calculating $\hat{EBP}$ based on only investment-grade or high-yield bonds does not dramatically influence its forecast power, suggesting our results are not specific to a certain segment of the corporate bond market. Moreover, our three main findings are similar when using the \citet{bkmx:21} metatopic aggregation rather than the GPT-derived metatopic aggregation we use. However, as that paper discusses, whether our use of topic-based categorization introduces look-ahead bias is an area in need of additional exploration.

Our main contribution is to study the news topics and sentiment that underpin the predictive power of the excess bond premium for macroeconomic fluctuations. Thus, it relates to a strand of literature on the link between the credit market and business cycle. \cite{philippon2009q} provides a theoretical foundation as to why credit spreads are related to macroeconomics through investments. The subsequent research (e.g. \citealt{mueller2008, gilchrist2009credit, GZ:12, culp2018option, krishnamurthy2020credit, benrephael2021flows, gilchrist2021term}) empirically investigates the link between credit spreads and macroeconomic dynamics. This strand often emphasizes the role of financial frictions in linking the credit cycle to the macroeconomy. In contrast, several papers explain the level and variation of credit spreads based on the investor's expectation of future economic growth (e.g., \citealt{BKS:2009, Chen:2010, GOURIO:2013}). Other literature includes the one that examines the firm-level growth predictability (e.g., \citealt{gilchrist2014firm}) and the study on international credit cycles (e.g., \citealt{bleaney2016europe}). \citet{SSZ:17} emphasize the role of credit market sentiment in predicting the business cycle. Our approach provides an empirical playing field that provides clear support for financial frictions and some evidence for sentiment in explaining the link between the credit cycle and macroeconomy, while highlighting potentially additional channels to explore.

This paper also contributes to the fast-growing literature on applying language models to improve our understanding of the macroeconomy and financial markets. \citet{nvix} provides compelling evidence that front-page news articles from the Wall Street Journal possess predictive power over stock volatility and returns. We employ the spirit of their analysis in using text to extend EBP backwards through time. \citet{AMJ:22} show how attention to macroeconomic news relates to announcement premia. \cite{manela} adapt the Multinomial Inverse Regression (MNIR) method, originally proposed by~\cite{mnir}, to build an efficient text selection process for financial analysis. \cite{color-finance-words} apply MNIR in the analysis of earnings calls to predict post-announcement returns. \cite{text-return-pred} construct a sentiment score specifically tailored to return prediction. More recently, \cite{llm-ret} demonstrate the superiority of word embeddings obtained from transformer models like BERT and RoBERTa over previous sentiment models and other embedding methods such as word2vec for downstream tasks like sentiment analysis and return prediction. 

Most closely related is \cite{bkmx:21}, who present a novel approach to assessing the state of the economy through textual analysis of business news. The authors use a topic model to distill Wall Street Journal articles from 1984 to 2017 into interpretable topics and measure the proportion of attention each topic receives over time. Their findings show that news attention aligns closely with various economic activities and can forecast aggregate stock market returns. Notably, the application of group lasso reveals that ``recession" is the topic with the highest predictive power for macroeconomic indicators, emphasizing its link to economic uncertainty. Our work advances this line of research by diving into the EBP and its specific forecast properties given its central role in understanding macroeconomic fluctuations.

Our work has implications for future research and is useful for policymakers and practitioners. First, while our results support the thesis that EBP predicts macroeconomic movements due to fluctuations in financial intermediary health, they also suggest the need for an expanded set of explanations related to politics, national policies, and potentially sentiment. Second, there are potential gains for practitioners and policymakers in filtering out non-predictive variation from EBP for the purposes of forecasting the macroeconomy and focusing only on the predictive variation driven by specific news topics. Finally, our results demonstrate that news attention is a main source of predictive variation for macroeconomic and financial time series that researchers and practitioners alike should devote effort to understanding in future research.

\section{Motivation and Methodology}\label{sec:methodology}
\subsection{Motivation}

In an influential paper, \citet{GZ:12} show that the excess bond premium, equal to the difference between the bottom-up average US corporate credit bond spread minus the averaged spread predicted from a default risk model, predicts macroeconomic movements from 1973 onward. They suggest that increases in EBP correspond to decreases in financial intermediary risk-bearing capacity which subsequently portend macroeconomic downturns. This explanation appeals to theories of financial frictions that link the health of the financial sector with the credit cycle and macroeconomic booms and busts (e.g., \citealp{BS:14}; \citealp{HK:13}; and the longstanding literature reviewed therein) and the extensive empirical evidence supporting such a link (e.g., \citealp{AS:10}; \citealp{HKM:17}).

On the other hand, the EBP is a residual of market prices from a measure of fundamentals, so its interpretation remains an open question. For example, \citet{SSZ:17} links the EBP to investor sentiment in addition to financial sector frictions. These sentiment-based links appeal to a broad class of theories that link biased expectations formation with booms and busts \citep{BGS:18,BGS:22,BGPS:24}.

Our approach is empirical and relies on decomposing variation in EBP into variation attributable to how much the news pays attention to different topics. \citet{bkmx:21} show that variation in the proportion of news attention paid to 180 different topics can track economic activity and forecast macroeconomic dynamics. We build on their work by focusing on quantifying which topics drive the EBP and its forecast power for the macroeconomy. Doing so offers empirical guidance on the relative importance of mechanisms such as frictions in financial intermediation versus sentiment, acknowledging that the truth is likely a mix of these explanations. Moreover, following \citet{nvix}, we can extend EBP backwards through time to evaluate EBP's forecast power going back to the start of the 20th century.

\subsection{Data sources and method}

We start by retrieving the EBP and standard macroeconomic time series from the websites of the Federal Reserve and St. Louis Federal Reserve FRED system.\footnote{EBP source: \url{https://www.federalreserve.gov/econres/notes/feds-notes/updating-the-recession-risk-and-the-excess-bond-premium-20161006.html}.} For the long-term analysis going back to the early 20th century, we obtain macroeconomic data from the Global Financial Data (GFD), including the consumer price index, unemployment rate, industrial production, gross domestic product, and fixed investment. GFD provides longer time series than FRED because it collects data from multiple historical sources.  

As in \citet{manela}, we retrieve front-page Wall Street Journal news articles from July 1889 to 2023 from ProQuest TDM Studio, a cloud-based tool that allows for text, data mining, and licensed extraction of text. We pre-process this data using standard techniques that we describe in Appendix \ref{app:technical}.

Figure \ref{fig:summary-stats} Panel A shows the average number of daily WSJ articles per month since July 1982. Prior to 1940, the number of articles varies greatly as the WSJ front page format changed over time. After 1940, the format remained the same and the average number of daily news articles is about eight.

Our main variable of interest, $\hat{EBP}$, equals a projection of EBP onto topic-level measures of news attention across the 180-topic model of \citet{bkmx:21}. We use a lasso regression to predict our model, which effectively selects only relevant topics and estimates the attention weights that EBP pays to the measured attention in each topic.

Specifically, for a given month $t$ and topic $k$, the estimated attention that news articles in month $t$ allocate to topic $k$ is determined by the frequency of terms associated with $k$:
\begin{align*}
    \hat{\theta}_{k,t} = \frac{\sum_{i=1}^{N_t}\mathbbm{1}(\hat{z}_{i,t}=k)}{\sum_{q=1}^K\sum_{i=1}^{N_t}\mathbbm{1}(\hat{z}_{i,t}=q)},
\end{align*}
where $\hat{z}_{i,t}$ represents topic assignment of word $i$ in the total vocabulary for month $t$, $K=180$ represents the total number of topics, and $N_t$ is the total vocabulary count in all articles in month $t$.  Note that we pool together all articles in a month.

We then fit $EBP_t$ to $\{ \hat{\theta}_{k,t} \}$ using a lasso procedure. Our estimated $\hat{EBP}$ equals:
\begin{align}\label{eq:ebp}
    \hat{EBP}_t = c + \sum_{k \in K^\ast} {w \cdot  \hat{\theta}_{k,t}},
\end{align}
where $c$ is a constant term, $K^\ast$ represents the topics selected by the lasso algorithm, and $w$ represents the estimated weights. Appendix \ref{app:technical} contains details.

We fit the model first for a training period covering 1973-1988 and then use an expanding window to update the model each month afterward. As we discuss below, our main analysis uses a conservative ``out-of-sample'' expanding-window approach where we lag the weights by one period so that $w=w_{k,t-1}$ in Equation \ref{eq:ebp} outside of the training sample. Figure \ref{fig:summary-stats} Panel B shows the number of topics selected over time, with the number of topics receiving a positive weight and negative weight. Our process typically selects about two-thirds of the possible topics, with an equal split between topics with positive and negative weights.

\subsection{How well does $\hat{EBP}$ fit $EBP$?}

Table \ref{tab:EBP-fit} reports diagnostic results from a regression of:
\begin{equation*}
    EBP_t = \alpha + \beta \hat{EBP}_t + \varepsilon_t.
\end{equation*}
A coefficient of $\beta = 1$ and $\alpha = 0$ would indicate that $\hat{EBP}$ is an unbiased estimate of $EBP$. Large values of $R^2$ and small root-mean-squared-errors and mean-absolute-errors would indicate that our estimates explain a significant share of variance with small average errors.

Column 1 reports that $\beta = 1.12$ with $\alpha=-0.01$ in our training period from 1973-1987 with a $R^2$ value of 86\%. After 1987, we use an expanding window approach to re-estimate the model. In columns 2 and 3, we re-estimate the new model contemporaneously with all data through month $t$ constructing $\hat{EBP}_t$; that is, we use $w = w_{k,t}$ in Equation \ref{eq:ebp}. We label this approach an ``in-sample'' estimation approach. Over the entire 1973-2023 sample, our estimate of $\hat{EBP}$ predicts $EBP$ with $\beta=1.01$ and $\alpha=0.00$ and an $R^2$ of 79\% (column 2). Even when we focus on the post-training sample period from 1988-2023, the values are very close: $\beta=1.00$, $\alpha=0.00$, and $R^2=78\%$ (column 3).

In Column 4, we take a more conservative ``out-of-sample'' approach to updating the model after the training period by delaying re-estimation by one month. Specifically, we feed the new text from month $t$ into a stale model estimated through month $t\mbox{-}1$ to construct $\hat{EBP}_t$ and apply $w=w_{k,t-1}$ in Equation \ref{eq:ebp}. We view this approach as conservative from a forecasting perspective as it will introduce noise into our estimates of $\hat{EBP}_t$, particularly at the start of the post-training period. The results show that $\hat{EBP}_t$ nevertheless predicts $EBP_t$ with $\beta=0.73$, $\alpha=0.01$, and an $R^2$ of 45\%.

Figure \ref{fig:ebp-fitting} illustrates these results by plotting the true values of $EBP$, the ``in-sample'' construction of $\hat{EBP}$, and the ``out-of-sample'' construction of $\hat{EBP}$. Consistent with the discussion above, the ``out-of-sample'' approach is noisier due to its conservative approach, yet delivers an overall fairly consistent fit with $EBP$.

\section{Understanding $\hat{EBP}$}\label{sec:understandingebp}
\subsection{$\hat{EBP}$ forecasts macroeconomic fluctuations}

As a validation exercise, we first show that $\hat{EBP}$ predicts movements in macroeconomic quantities. We follow \citet{GZ:12} and forecast the $h$-period change in monthly private nonfarm payroll employment, the unemployment rate, and industrial production using the following specification:
\begin{align}\label{eq:predict}
\nabla^h Y_{t+h} &= \alpha + \sum_{i=1}^{p=3} \beta_i \nabla Y_{t-i} + \gamma_1 TS_t + \gamma_2 RFF_t + \gamma_3 GZF_t + \gamma_4 \hat{EBP}_t + \gamma_5 (\hat{EBP}\mbox{-}RES_t) + \varepsilon_{t+h}.
\end{align}
where $\nabla^h Y_{t+h} \equiv \frac{c}{h+1} \ln \left( \frac{Y_{t+h}}{Y_{t-1}} \right)$, $c$ represents an annualization constant (1200 for monthly data and 400 for quarterly), $TS_t$ represents the 10-year minus 3-month Treasury yield term spread, $RFF_t$ represents the real federal funds rate, $GZF_t$ represents the default-risk-implied credit spread, $\hat{EBP}_t$ represents the predicted value of $EBP$ from the news at time $t$, and $\hat{EBP}\mbox{-}RES_t = EBP_t - \hat{EBP}_t$ represents the difference of the true $EBP_t$ and text-predicted value $\hat{EBP}_t$.

Our key variable of interest is $\hat{EBP}$, which we stress that we construct by following the conservative ``out-of-sample'' expanding-window approach where we feed month-$t$ text as input into a stale model estimated in month $t\mbox{-}1$. As Table \ref{tab:EBP-fit} column 4 highlights, this approach conservatively leaves a significant amount of variation in $EBP_t$ as unexplained when constructing $\hat{EBP}_t$ this way compared to alternative methods. We estimate Equation \ref{eq:predict} using all available time series to reliably assess whether there is a relationship between economic growth and the news attention available to the agent at any point in the past. 

In Equation \ref{eq:predict}, we include $TS_t$ and $RFF_t$ as control variables following \citet{GZ:12} as they are well-known predictors of the macroeconomy. We also follow their work in including $GZF_t$. We include $\hat{EBP}\mbox{-}RES_t$ to test whether $\hat{EBP}_t$ soaks up all of the variation in $EBP_t$ that predicts macroeconomic movements.\footnote{We note the following other features of Equation \ref{eq:predict} where we follow \citet{GZ:12}. First, we do not apply the log transform when studying the unemployment rate and calculate arithmetic differences instead. Second, we allow for nowcasting and the possibility that $Y_{t}$ is not yet observed at the time of the forecast date by placing $Y_{t-1}$ in the denominator of $\nabla^h Y_{t+h}$ instead of $Y_{t}$. We depart from \citet{GZ:12} in using a fixed lag structure of 3 months instead of an AIC-determined lag structure for simplicity. Appendix \ref{app:technical} contains additional details about variables.} Note that $GZF_t+\hat{EBP}_t+\hat{EBP}\mbox{-}RES_t = GZF_t + EBP_t$ which equals the non-decomposed average credit spread.

Table \ref{tab:prediction-macro-fluctuations} Panel A, columns 1-3 report the estimates of Equation \ref{eq:predict} for forecasting at the $h=3$ horizon. The estimates show that high levels of $\hat{EBP}$ forecast 3-month decreases in payrolls, increases in unemployment, and decreases in industrial production. The estimated coefficients are highly economically significant: a one percentage point increase in $\hat{EBP}$ predicts a 2.74\% annualized decrease in nonfarm payrolls, a 1.81 percentage point annualized increase in the unemployment rate, and a 6.82\% annualized decrease in industrial production.  The estimates also show that $\hat{EBP}$ captures a significant amount of predictive variation for macroeconomic movements. Specifically, the last three rows of the table reports \citet{SHAP}'s SHAP (SHaply Additive exPlanations) value, which measures the difference between the prediction model's output variation with and without each explanatory variable.\footnote{We provide the details of the SHAP calculation in Section \ref{subsec:SHAP} in the Appendix.} For $\hat{EBP}$, the SHAP value is significantly higher than that for $GZF$, supporting the economic significance of $EBP$.  However, $\hat{EBP}\mbox{-}RES$ predicts macroeconomic movements as well, suggesting that $\hat{EBP}$ does not capture all of the variation in $EBP$ that predicts the macroeconomy.

Columns 4-5 report additional results. Column 4 reports estimates of Equation \ref{eq:predict} where we predict quarterly log GDP changes for $h=1$ after controlling for its lagged values (including 4 lags), while column 5 reports estimates from a probit forecasting regression where the dependent variable is an indicator for whether the economy is in recession in month $t+h$, as dated by the National Bureau of Economic Research (NBER). In both columns, we reach similar conclusions: $\hat{EBP}$ is an important variable in forecasting the macroeconomy as in columns 1-3.

Table \ref{tab:prediction-macro-fluctuations} Panel B reports estimates that include $EBP$ rather than $\hat{EBP}$ and $\hat{EBP}\mbox{-}RES_t$ as right-hand-side variables for comparison purposes. The point estimates for $\hat{EBP}$ in Panel A are within one standard error of those for $EBP$ in Panel B for almost all columns. For example, in Panel A, a one percentage point increase in $\hat{EBP}$ predicts a 2.74\% decline in annualized payroll growth.  In Panel B, the same increase in $EBP$ predicts a similar decline of 2.19\%.  SHAP values for $\hat{EBP}$ and $EBP$ suggest that the two variables are comparable in economic importance in predicting macroeconomic movements, although values for $EBP$ are larger.

To illustrate the importance of $\hat{EBP}$, Figure \ref{fig:recession-prob} plots predicted recession probabilities when using $EBP$ as a predictor alongside predicted probabilities when using $\hat{EBP}$ but excluding $\hat{EBP}\mbox{-}RES$. The graph shows that the two predicted probabilities align closely even though the latter set of predicted probabilities does not include variation from $\hat{EBP}\mbox{-}RES$.

Overall, $\hat{EBP}$ captures significant variation in $EBP$ that predicts macroeconomic quantities at the 3-month horizon. Appendix Table \ref{tab:prediction-macro-fluctuations-12m} shows that similar insights hold when studying $h=12$-month or $h=4$-quarter horizons.

\subsection{Topic analysis}
To better understand why $EBP$ varies, in this section we dissect which underlying topics are associated with an increase and a decrease in $\hat{EBP}.$

\subsubsection{Which topics move $\hat{EBP}$ through news attention?}

Table \ref{tab:topic-attention} reports the aggregated attention weights and explained variance of various news topics for $\hat{EBP}$ (Panel A) and $\hat{GZF}$ (Panel B), where the latter equals the fitted values of the default-risk-implied spread $GZF_t$ following the exact same methodology we applied to $EBP_t$. For ease of interpretation, we aggregate the 180 news topics of \citet{bkmx:21} into ten metatopic categories using the GPT-o1 large language model (LLM). We explain this aggregation process in more detail in Appendix \ref{sec:gptmeta} and Appendix Table \ref{tab:metatopic-classification}.

The top contributors of explained variance that drive up $\hat{EBP}$ include crises-related topics (both financial and non-financial), negative sentiment, and financial intermediaries, while top contributors that drive down $\hat{EBP}$ include politics and industry. By way of comparison, the top topics that drive up $\hat{GZF}$ are general news about financial markets and the economy in addition to crises and financial intermediaries, and the top topics that drive it down are also politics and industry. A category of ``other'' topics (listed in Appendix Table \ref{tab:metatopic-classification}) also drives down $\hat{EBP}$, and we show later that variation in this topic does not flow through to forecast power for macroeconomic movements.

The two panels show that $\hat{EBP}$ and $\hat{GZF}$ behave differently in a few important respects despite these overall similar patterns. $\hat{EBP}$ is comparatively more sensitive to news about crises and financial intermediaries than $\hat{GZF}$ due to its much larger raw weight. Political news has a much greater variance share and drives down $\hat{EBP}$ compared to $\hat{GZF}$, where it is roughly neutral. Finally, general news about financial markets and the economy drives down $\hat{EBP}$ but drives up $\hat{GZF}$, with a far smaller variance share of the former. 

These attention weights may vary over time, as the technological and political environment changes over the long term. Therefore, the type of topic that is important in today's economy may be different from what was important in the 1980s.  To this end, we estimate the fit of the EBP on a rolling window basis and obtain the time-varying weight on each metatopic.  Figure \ref{fig:metatopic-time} shows these time-varying attention weights.  We find that there is an interesting variation in the weights for EBP (Panel A).  For example, the role of positive sentiment (which decreases EBP) and financial intermediaries (which increases EBP) has been relatively stable since the 1980s.  However, the weight of ``financial crisis'' increases significantly since 2008, while the weight of ``politics'' changes from positive in the 1980s to negative in the 2010s. 

To provide additional narrative evidence, we highlight representative articles in the financial intermediaries, financial crisis, politics, and crisis/disasters metatopics. For each metatopic $M$, we select the month $t$ from January 1988 to December 2023 when $\hat{\text{EBP}}\text{-M}$ has the largest magnitude. Within this month, we choose the article that garnered the most attention to the metatopic $M$. Table \ref{tab:representative-articles} lists the top 3 articles for each topic in the selected month, along with the associated $\hat{\text{EBP}}\text{-M}$. 

For financial intermediaries, the top articles are from December 1988. While one article discusses Drexel Burnham Lambert Inc., which went bankrupt in 1990, another article is about Morgan Stanley employees participating in an Outward Bound program in New York City. In contrast, for politics, the top articles are from September 2012.  In that month, the most representative articles were about then-presidential candidate Mitt Romney's tax disclosures (as well as several articles about the election more broadly not in the table) and Israel-Iran-U.S. geopolitics. This disparity in content illustrates that the key variation in our model comes from attention to topics, rather than the content itself.

\subsubsection{Which topics predict macro movements through $\hat{EBP}$?}

We next study which metatopics forecast movements in the macroeconomy through associated movements in $\hat{EBP}$. For every metatopic $M$, we take all topics $k \in M$ and calculate the component of $\hat{EBP}$ associated specifically with that metatopic, $\hat{EBP}\mbox{-}[M]_t \equiv \sum_k{w_{k,t-1} \theta_{k,t}}$, where $w_{k,t-1}$ and $\theta_{k,t}$ equal the (stale) model weight and news attention for topic $k$, respectively. Note that $\hat{EBP}_t = \alpha_{t-1} + \sum_M \hat{EBP}\mbox{-}[M]_t$ for the constant term $\alpha_{t-1}$ from the text model. Figure \ref{fig:topic-decomposition} illustrates a few examples such time series. Table \ref{tab:topic-decomp-prediction-gpt} presents regression estimates analogous to the forecasting exercise in Equation \ref{eq:predict} but where we substitute the set of $\{\hat{EBP}\mbox{-}[M]_t\}$ for $\hat{EBP}$.

The estimates suggest that the variation in $\hat{EBP}$ induced by news attention to different metatopics flows through to forecast power for macroeconomic movements, but not necessarily for all topics. For example, greater news attention to topics related to financial crises, negative sentiment, and financial institutions tend to portend economic downturns as suggested by the statistically significant loadings on $\hat{EBP}\mbox{-FCRIS}$, $\hat{EBP}\mbox{-NEGSENT}$, and $\hat{EBP}\mbox{-FI}$. Similarly, greater attention to politics, by decreasing values of $\hat{EBP}\mbox{-PLTC}$, portends upturns. However, $\hat{EBP}\mbox{-CRISDIS}$ and $\hat{EBP}\mbox{-IND}$ do not have forecast power for the macroeconomy even though news attention to non-financial crises/disasters and industry are important components of variation in $\hat{EBP}$ from Table \ref{tab:topic-attention}. Moreover, the greater adjusted $R^2$ values for predicting IPM and recession risk in Table \ref{tab:topic-decomp-prediction-gpt} versus those in Table \ref{tab:prediction-macro-fluctuations} Panel B indicate significant gains in explanatory power from using the meta-topic decomposition of $\hat{EBP}$ compared to using ``ground-truth'' values of $EBP$.

\subsection{Sentiment analysis}

The previous analysis suggests that news attention to topics reasonably associated with negative sentiment contributes to variation in $\hat{EBP}$ and forecasts macroeconomic movements. Sentiment is interesting because it has been debated in the literature (e.g., \citealt{culp2018option}) whether EBP is driven by credit risk premiums demanded by rational agents for tail risk exposures or by sentiment decoupled from fundamentals.  Moreover, our sentiment measure differs from what is proposed in the literature as it is based on attention to sentiment-related topics.  Thus, for comparison, we next construct three empirical measures that directly incorporate information about sentiment. We summarize their construction below and provide details in Appendix \ref{app:technical}.

Our first measure, $SENT^{LM}$, measures monthly aggregate market sentiment and equals the polar sentiment score of all words in a month, $SENT^{LM}_t \equiv \frac{c_t^+ - c_t^-}{c_t^+ + c_t^-}$. Here, $c_t^+$ and $c_t^-$ represent the number of positive and negative sentiment words each month from the \citet{LM:11} dictionary.

Our second measure, $SENT^{A}$, is a topic-weighted sentiment measure equal to the weighted average polar sentiment score across news topics. Specifically, we calculate $SENT^A \equiv \sum_k w_{k,t-1} s_{k,t}$ for topic-level sentiment scores $s_{t,k}$. We construct $s_{t,k}$ by taking the weighted-sum of article-level sentiment scores $s_a$ with weights $\theta_{a,k}$ equal to the attention share of each article $a$ to topic $k$: $s_{t,k} \equiv \sum_a s_a \theta_{a,k}$.

Our third measure, $\hat{\hat{EBP}}$, equals the fitted value of $EBP$ to $SENT^{LM}$ using the same out-of-sample, expanding-window approach we use when constructing $\hat{EBP}$: We fit the model in the training sample and then calculate $\hat{\hat{EBP}}_t$ from 1988-onwards by feeding time-$t$ news text into a stale model estimated through time-$t-1$ data. 

To help build intuition, Figure \ref{fig:sentiment} plots each of these three sentiment measures. Panel A plots $SENT^{LM}$ and $SENT^A$ while Panel B plots $\hat{\hat{EBP}}$, including $EBP$ and $\hat{EBP}$ for reference. $SENT^{LM}$ and $SENT^A$ are correlated, although $SENT^A$ is visually more responsive to events like the 2008 financial crisis. Because $SENT^{LM}$ appears to be less sensitive to recessions, the projection of EBP on it ($\hat{\hat{EBP}}$) differs significantly from $\hat{EBP}$ although there is a modest correlation of 0.31.

Table \ref{tab:sentiment} reports estimates from the predictive regressions in Equation \ref{eq:predict} and results in Table \ref{tab:prediction-macro-fluctuations} that replaces $\hat{EBP}$, $GZF$, and $\hat{EBP}\mbox{-}RES$ on the right-hand-side with the sentiment measures above to forecast non-farm payrolls, unemployment, and industrial production.  For the specification including $\hat{\hat{EBP}}$, we include the difference between EBP and its sentiment-driven part denoted $\hat{\hat{EBP}}\mbox{-}RES$, for comparison. Columns 1-6 show that sentiment proxies, $SENT^{LM}$ and $SENT^{A}$ have positive forecast power for macroeconomic movements. 
Columns 7-9 show that $\hat{\hat{EBP}}$ -- the component of $EBP$ associated with sentiment measures -- also predicts macroeconomic movements. However, the $R^2$ values in columns 1-6 are much lower than the comparable columns in Table \ref{tab:prediction-macro-fluctuations}. Moreover, SHAP values for $\hat{\hat{EBP}}$ are all lower than those for $\hat{\hat{EBP}}\mbox{-}RES$ in columns 7-9, which is the opposite pattern for $\hat{EBP}$ in Table \ref{tab:prediction-macro-fluctuations}. Thus, while our sentiment measures have some power to forecast macroeconomic movements, $EBP$ and the topic-driven $\hat{EBP}$ may potentially be more economically and statistically important predictors.

We next run a horse race and estimate a multivariate regression version of Equation \ref{eq:predict} using both the topic-based predictor of EBP $\hat{EBP}$ and sentiment proxies. Table \ref{tab:sentiment-attention} reports results that show that $\hat{EBP}$ drives out the forecast power of our sentiment variables when including both sets of variables as predictors. Columns 1-6 show that $SENT^A$ and $SENT^{LM}$ retain at best marginal predictive power. Columns 7-9 show that $\hat{\hat{EBP}}$ does provide some incremental forecast power even when including $\hat{EBP}$. However, across all specifications, the SHAP values for sentiment measures are much lower than in Table \ref{tab:sentiment}, and much lower than $\hat{EBP}$.

Overall, our results show that measures of news attention, rather than sentiment, capture the economically important variation in $EBP$ that forecasts macroeconomic movements. While our results do not preclude a role for sentiment in driving EBP or macroeconomic movements, any such role flows through our attention-based measures rather than measures based directly on sentiment-charged words.

\section{Historical Analysis}\label{sec:historical}

Values of EBP in \cite{GZ:12} start in 1973 due to the availability of security-level credit spread data. The advantage of topic-based prediction of EBP is that the news articles are available from the late 19th century. Therefore, we are able to construct our topic-based EBP by extending the original sample by almost a century.  On the other hand, this requires us to assume that the relationship between credit spreads and the topics that market participants pay attention to is stable over a long period of time.  We validate this assumption by establishing the link between the extrapolated series of $\hat{EBP}$ and macro variables prior to 1973.

We construct the historical $\hat{EBP}$ from July 1889 to December 1972 (out-of-sample prediction) using the period of January 1973 to December 2023 as the in-sample training period. We then evaluate its forecast power for changes in CPI starting from 1915, industrial production (1915), unemployment (1929), GDP (1947), and non-residential fixed investment (1947). The starting point of each data series determines the start date of our data window.

Figure \ref{fig:backward-projection} plots our estimated series $\hat{EBP}$ both during the in-sample training period from 1973-present and for the out-of-sample fit from 1889. Visually, there are periods when elevated levels of $\hat{EBP}$ tend to portend recessionary periods, although the overall picture is not conclusive likely due to noise. Specifically, we expect the series to contain some noise as our topic categorization may become less relevant further back in history. 

For a more systematic analysis, Table \ref{tab:backward-projection} reports results from the following forecast regression:
\begin{align}\label{eq:histforecast}
    \nabla^h Y_{t+h} &= \alpha + \sum_{i=1}^{p=3} \beta_i \nabla Y_{t-i} + \gamma_1 \hat{\text{EBP}}_t + \gamma_2 \text{BAA-Yield}_t + \gamma_3 \text{TBill-Yield}_t + \varepsilon_{t+h},
\end{align}
where Panel A reports results for monthly frequency variables (changes in CPI, unemployment, and industrial production) while Panel B does so for quarterly frequency variables (changes in GDP and NFI) with 4 lags. We include the Moody's Baa corporate bond yield and short-term T-Bill rates to control for the predictive power of credit spreads and interest rates for the macroeconomy. The specification is more parsimonious than that of Equation \ref{eq:predict} due to data limitations.  We predict over both 3-month and 12-month horizons.  To prevent our in-sample fit to EBP after 1973 from predicting the outcome, we use the sample before 1972 for this analysis.

The results are broadly in line with Table \ref{tab:prediction-macro-fluctuations} in showing that elevated levels of $\hat{EBP}$ predict macroeconomic downturns such as GDP growth and unemployment. For example, a one percentage point increase in $\hat{EBP}$ predicts a 4.05 percentage point (annualized) decline in GDP growth three months later and a 2.68 percentage point increase in the unemployment rate.  The magnitude of these estimates is larger than the estimates from the recent sample reported in Table \ref{tab:prediction-macro-fluctuations}, likely because the historical sample includes the Great Depression period of the 1930s. Higher $\hat{EBP}$ also predicts deflation and weakening non-residential fixed investment. However, $\hat{EBP}$ does not have statistically significant predictive power for industrial production in the historical time series.  Still, the sign is consistent with greater values predicting lower production.

In Appendix Tables \ref{tab:historical-topic-decomp-gpt} and \ref{tab:historical-topic-decomp-gptq}, we repeat the forecasting regressions by topic, mimicking Table \ref{tab:topic-decomp-prediction-gpt} but using historical data before 1973.  Comparing the loading on $\hat{EBP}$ by topics, we find that politics-driven $\hat{EBP}$, $\hat{\text{EBP}}$-PLTC, retains signs consistent with the main sample.  For example, a politics-driven rise in $\hat{EBP}$ forecasts a significant decline in GDP and NFI in one and four quarters.  The consistency in the link between topics and economic growth underscores the soundness of this sample extension.

To understand why $\hat{EBP}$ predicts economic growth rates despite its estimation noise, we examine historical episodes related to corporate credit.  For $\hat{EBP}$ to predict an economic downturn, it must rise before the event.  Table \ref{tab:ebp-historical-crisis} reports average values of $\hat{EBP}$ in the 12 months before the start of historical banking panics and NBER recessions. We date our list of banking panics using data from \citet{BD:20}. Such panics tend to be extreme events that have severe economic consequences, although they are not necessary for banking crises \citep{BVX:21}. We find that the average values of $\hat{EBP}$ prior to these events range from 0.076\% (1984) to 0.635\% (1890) and are typically substantially greater than values of $\hat{EBP}$ in all other months (0.040\%).  In addition, $\hat{EBP}$ before the NBER recession averages 0.149 percentage points, which is also higher than the average of the other months.  Taken together, the results in Table \ref{tab:ebp-historical-crisis} provide narrative evidence consistent with the regression results above.

\section{Robustness}\label{sec:robust}
In this section, we extend our baseline results using alternative methods to create $\hat{EBP}$ and classify the topics into metatopics.

\subsection{Investment-grade versus high-yield debt}

The existing literature on corporate credit (e.g. \citealt{Greenwood-Hanson-2013}) suggests that high-yield (HY) bonds that have higher default risks behave differently from investment-grade (IG) bonds over the business cycle.  In particular, HY bonds are more illiquid and their yields are more sensitive to credit conditions than IG bonds. Moreover, demand for HY and IG bonds varies with different components of the business cycle.

To investigate the potential difference in information between IG bonds and HY bonds, we create rating-specific EBP series, $\hat{EBP}\mbox{-}IG$ and $\hat{EBP}\mbox{-}HY$, by calculating bond-level excess bond premiums and averaging them within IG and HY segments following the general process of \citet{GZ:12}.  We then project them onto the topics as we do in the main analysis and estimate the 3-month ahead forecasting regression in Equation \ref{eq:predict}.  

Table \ref{tab:ig-hy} shows the forecasting results using $\hat{EBP}\mbox{-}IG$ (Panel A) and $\hat{EBP}\mbox{-}HY$ (Panel B). The estimated coefficients and $R^2$ values suggest that the two series yield similar degrees of forecast power for the macroeconomy. For example, the $R^2$ using the IG-based $\hat{EBP}$ ranges from 0.15 to 0.38, while that using the HY-based $\hat{EBP}$ ranges from 0.15 to 0.39. SHAP values do suggest that the relative importance of $\hat{EBP}\mbox{-}IG$ versus other variables is greater in Panel A than the relative importance of $\hat{EBP}\mbox{-}HY$ in Panel B. Overall, however, the topic-based projection of EBP predicts economic growth regardless of the default risk of the bond issuers.

\subsection{Alternative metatopic classifications}
We provide insights into the topic-based drivers of EBP by classifying the 180 original topics into `interpretable' metatopics using GPT-o1.  In this section, we show that the key insights from our analysis are not sensitive to this choice of metatopics.  

To this end, instead of relying on our meta-classification, we use the metatopic classification of \citet{bkmx:21}, or ``BKMX classification.'' We list this topic aggregation in Appendix Table \ref{tab:metatopic-classification-bkmx}.

The BKMX classification differs from our GPT-driven classification in a few key ways. For example, ``announcements'' form a separate metatopic in the former classification, whereas the GPT classification divides up this metatopic among many others. We also had GPT attempt to divide topics into positive and negative sentiment topics, which the BKMX classification avoids by design. We do not view any particular classification as superior ex-ante but rather approach this exercise with the goal of looking for common threads in the analysis.

Table \ref{tab:topic-attention-bkmx} and Figure \ref{fig:metatopic-time-bkmx} report the metatopic-based decomposition of EBP using the BKMX classification.  
Overall, several similarities emerge. News about financial crises and intermediaries are top contributors to the explained variance that drives up $\hat{EBP}$. Indeed, the financial intermediaries topic receives a large raw weight that is consistent through time (see Figure \ref{fig:metatopic-time-bkmx}). Moreover, the role of crises and intermediaries is more pronounced for $\hat{EBP}$ than for the default-risk-implied component of spreads, $\hat{GZF}$. News about national policies is also a top contributor to variance and drives down $\hat{EBP}$, similar to the role of the politics metatopic from our GPT categorization.

Nonetheless, there are some notable distinctions between the two methods. In the BKMX classification, national policies also play an important role in explaining the variance of $\hat{GZF}$, whereas ``politics'' is roughly neutral in our GPT analysis. Moreover, in the BKMX classification, ``announcements'' are a top contributor to the variance of both $\hat{EBP}$ and $\hat{GZF}$, which is not a metatopic in our main results.

Table \ref{tab:topic-decomp-prediction-bkmx} report results from an exercise similar to that of Table \ref{tab:topic-decomp-prediction-gpt} studying whether metatopic-level components of $\hat{EBP}$ predict macroeconomic movements, this time using the BKMX classification. The results confirm that the results from our GPT analysis that news attention to financial institutions and politics forecasts these movements through their associated metatopic-level component of $\hat{EBP}$.

\section{Conclusion}\label{sec:conclusion}

Our results highlight that financial frictions are an important source of variation for the EBP that predicts the macroeconomy, but not the only source. Sentiment likely plays a role, although the evidence is mixed insofar as our attention-based reconstruction of EBP drives out direct measures of charged sentiment when predicting the macroeconomy. To the extent that sentiment plays a role in explaining why EBP predicts the macroeconomy, its role empirically flows through attention-based measures. Moreover, additional news topics appear to be important drivers of EBP's predictive variation. News attention to politics and national policies in particular appear to drive down the EBP and portend increased growth for the economy.

Our work has several implications for practice and future research. For practice, policymakers and practitioners can use attention to news topics to filter EBP for non-predictive sources of variation and extract the most predictive sources for macroeconomic fluctuations. Our results suggest that they should focus on attention-based measures of news rather than direct measures of charged sentiment, which many data vendors calculate. For future research, our results suggest the need for additional research into an expanded set of explanations for what might drive EBP and macroeconomic fluctuations beyond financial intermediary frictions: such a set should include politics, national policies, and to some extent, sentiment. Moreover, our results highlight the promise of using text-based analysis to better understand how to interpret important yet anomalous financial and macroeconomic time series. Future research should further build on these methods.

\small
\clearpage
\setstretch{1.1}
\bibliography{main}
\normalsize
\setstretch{1.5}

\input{tables}

\newpage
\clearpage

\small
\setstretch{1.1}
\input{appendix}

\end{document}

%% file: tables.tex
\clearpage
\newpage
\begin{table}[!htb]
\caption{Fit of the Topic Model to Excess Bond Premium}\label{tab:EBP-fit}
\small
{This table reports the coefficients and standard errors of the regression:
\begin{equation*}
    EBP_t = \alpha + \beta \hat{EBP}_t + \varepsilon_t,
\end{equation*}
where $EBP$ is the excess bond premium of \cite{GZ:12} and $\hat{EBP}$ is the EBP predicted by the topic model, estimated using in-sample (IS) and the out-of-sample (OOS) fitting methods.  The standard errors are adjusted for serial correlation with Newy-West $T^{1/4}$ lags. The end of the sample for fitting is 2023-12-01.
}

\bigskip
\centering
\resizebox*{\textwidth}{!}{\input{tables/table1.tex}}

\end{table}

\newpage
\clearpage
\begin{table}[!htb]
\caption{Predicting Macroeconomic Fluctuations}\label{tab:prediction-macro-fluctuations}
\small
{
Columns 1 to 3 of this table report the monthly regression coefficients and their standard errors, 
\begin{align*}
\nabla^h Y_{t+h} &= \alpha + \sum_{i=1}^{p=3} \beta_i \nabla Y_{t-i} + \gamma_1 TS_t + \gamma_2 RFF_t + \gamma_3 GZF_t + \gamma_4 \hat{EBP}_t + \gamma_5 (\hat{EBP}\mbox{-}RES_t) + \varepsilon_{t+h},
\end{align*}
where $GZF$ is the credit spreads explained by the fundamentals, $\hat{EBP}$ is the excess bond premium projected on the topic model, and $\hat{EBP}-RES$ is the excess bond premium unaccounted by the topic model. $TS$ is the term spread and $RFF$ is real federal funds rate.  EMP-SA is seasonally-adjusted private nonfarm payrolls, UER is the unemployment rate, IPM is industrial production.
Columns 4 and 5 report the quarterly regression of gross domestic product (GDP) and the monthly probit regression of NBER recession dummy (Recession), respectively:
\begin{align*}
\nabla^h Y_{t+h} &= \alpha + \sum_{i=1}^{p=4} \beta_i \nabla Y_{t-i} + \gamma_1 TS_t + \gamma_2 RFF_t + \gamma_3 GZF_t + \gamma_4 \hat{EBP}_t + \gamma_5 (\hat{EBP}\mbox{-}RES_t) + \varepsilon_{t+h},\\
P(NBER_{t:t+h}) &= \Phi(\alpha + \gamma_1 TS_t + \gamma_2 RFF_t + \gamma_3 GZF_t + \gamma_4 \hat{EBP}_t + \gamma_5 (\hat{EBP}\mbox{-}RES_t) + \varepsilon_{t+h}).
\end{align*}
We forecast macroeconomic variables 3 months ahead ($h=3$ for monthly and $h=1$ for quarterly variables) using $\hat{\text{EBP}}$ estimated out-of-sample. The standard errors account for the Newy-West three lags for EMP-SA, UER, and IPM, one lag for GDP, and 12 lags for Recession. $SHAP$ is the Shapley Additive Explanations value of \cite{SHAP} for each of the key explanatory variables. The end of sample for prediction is 2023-07-01 so that both the 3-month and 12-month forecast have the same sample size. Note that the sample of the macro variables (e.g. IPM) ends on 2024-07-01.
}

\bigskip
\centering

Panel A. $\hat{\text{EBP}}$\\
\resizebox*{1\textwidth}{!}{\input{tables/table2-ver2.tex}}

\end{table}

\clearpage
\newpage
\begin{table*}[!htb]
\centering
{
Table~\ref{tab:prediction-macro-fluctuations} continued.
}

Panel B. EBP\\
\resizebox*{1\textwidth}{!}{\input{tables/table2b.tex}}

\end{table*}

\clearpage
\newpage
\begin{table}[!htb]
\caption{Topics of Attention}\label{tab:topic-attention}
\small
{
    This table reports the aggregated weights and explained variance of metatopics for text-based excess bond premium, $\hat{\text{EBP}}$, and the projection of the fundamental component of the GZ spread on texts, $\hat{\text{GZF}}$ over the time window 1973-01 to 2023-12.  We aggregate the 180 topics of \cite{bkmx:21} into ten metatopics and report the aggregated explained variance and weights for each of the metatopics.
}

\bigskip
\centering

A. $\hat{\text{EBP}}$\\ \smallskip
\resizebox*{1\textwidth}{!}{\input{tables/table3_EBP_base_gpt.tex}}

\vspace{.3cm}
B. $\hat{\text{GZF}}$\\ \smallskip
\resizebox*{1\textwidth}{!}{\input{tables/table3_GZF_base_gpt.tex}}
\end{table}

\begin{landscape}
\begin{table}[!htb]

\caption{Representative Articles}\label{tab:representative-articles}

{\small This table lists the three articles that have the highest contribution to the metatopic in the month when the excess bond premiums projected on the metatopic have the highest absolute value.  We list the four metatopics that are the significant contributors to the time-series variation in $\hat{EBP}$}, including financial intermediaries, financial crisis, politics, and crisis/disasters.
\smallskip

\centering
\resizebox*{1.4\textwidth}{!}{\input{tables/representative_articles}}
\end{table}
\end{landscape}

\clearpage
\newpage
\begin{table}[!htb]
\caption{Predicting Macroeconomic Fluctuations with Topic Decomposition (GPT)}\label{tab:topic-decomp-prediction-gpt}
\small
{
This table reports the regression coefficients and standard errors of the three-month-ahead macroeconomic forecasts in Table \ref{tab:prediction-macro-fluctuations}, replacing $\hat{EBP}$ with the excess bond premiums driven by each of the 11 metatopics categorized by GPT-o1.  The standard errors account for the Newy-West three lags for EMP-SA, UER, and IPM, one lag for GDP, and 12 lags for Recession.
}

\bigskip
\centering

\resizebox*{1\textwidth}{!}{\input{tables/table4-ver2-gpt.tex}}

\end{table}

\clearpage
\begin{landscape}
\newpage
\begin{table}[!htb]
\caption{Forecasting Regression Based on Sentiment Measures}\label{tab:sentiment}
\small
{
This table reports the regression coefficients and standard errors of the forecasting regressions: 
\begin{align*}
    \nabla^h Y_{t+h} &= \alpha + \sum_{i=1}^{p=3} \beta_i \nabla Y_{t-i} + \gamma_1 TS_t + \gamma_2 RFF_t + \gamma_3 SENT + \varepsilon_{t+h},
\end{align*}
where $SENT$ is either $SENT^{LM}$, which is the sentiment measure based on \cite{LM:11} dictionaries using all words in a month, $SENT^A$, which is a topic weighted sentiment score, or $\hat{\hat{EBP}}$, which is the fitted value of $EBP$ using $SENT^{LM}$.  Columns 1 to 3 use $\text{SENT}^{LM}$, columns 4 to 6 use $\text{SENT}^A$, columns 7 to 9 use $\hat{\hat{\text{EBP}}}$ to forecast the three macroeconomic variables. All the forecasts are 3-month/1-quarter based on the out-of-sample fit of the right-hand-side variables. The standard errors are adjusted for the Newy-West three lags for the monthly and one lag for quarterly data.
}

\bigskip
\centering

\resizebox*{1.2\textwidth}{!}{\input{tables/table5.tex}}

\end{table}
\end{landscape}

\clearpage
\begin{landscape}
\begin{table}[!htb]
\caption{Sentiment vs Attention}\label{tab:sentiment-attention}
{
\small
This table reports the regression coefficients and standard errors of forecasting regressions: 
\begin{align*}
\nabla^h Y_{t+h} &= \alpha + \sum_{i=1}^{p=3} \beta_i \nabla Y_{t-i} + \gamma_1 TS_t + \gamma_2 RFF_t + \gamma_3 GZF_t
 + \gamma_4 \text{SENT} + \gamma_5 \hat{EBP}_t + \gamma_6 \hat{EBP}\mbox{-}RES_t + \varepsilon_{t+h}.
\end{align*}
Columns 1 to 3 report the results using the sentiment measure using all words in a month ($\text{SENT}^{LM}$), columns 4 to 6 use the topic-weighted sentiment score ($\text{SENT}^A$), and columns 7 to 9 use the fitted value of $EBP$ using $\text{SENT}^{LM}$ ($\hat{\hat{\text{EBP}}}$). All the forecasts are 3-month/1-quarter based on the out-of-sample fit of the right-hand-side variables. The standard errors are adjusted for Newy-West three lags for monthly and four lags for quarterly data.
}

\bigskip
\centering

\resizebox*{1\textwidth}{!}{\input{tables/table6.tex}}

\end{table}
\end{landscape}

\clearpage
\newpage
\begin{table}[!htb]
\caption{Long-Run Forecasts Using Text-Based Excess Bond Premiums}\label{tab:backward-projection}

{\small
     This table reports the coefficient and standard errors of the forecasting regressions of macroeconomic variables.  Specifically, Panel A reports the regression coefficients and standard errors of the monthly forecasting regressions:
    \begin{align*}
        \nabla^h Y_{t+h} &= \alpha + \sum_{i=1}^{p=3} \beta_i \nabla Y_{t-i} + \gamma_1 \hat{\text{EBP}}_t + \gamma_2 \text{BAA-Yield}_t + \gamma_3 \text{TBill-Yield}_t + \varepsilon_{t+h}.
    \end{align*}
    Panel B reports the regression coefficients and standard errors of the quarterly forecasting regressions:
    \begin{align*}
        \nabla^h Y_{t+h} &= \alpha + \sum_{i=1}^{p=4} \beta_i \nabla Y_{t-i} + \gamma_1 \hat{\text{EBP}}_t + \gamma_2 \text{BAA-Yield}_t + \gamma_3 \text{TBill-Yield}_t + \varepsilon_{t+h}.
    \end{align*}
    For TBill-Yield, we use the 90-day secondary market Treasury Bill rates. For each month, we used the yield on the last available day in the month. The monthly regression (Panel A) starts from 1915 for the consumer price index (CPI) and industrial production (IPM), 1929 for unemployment (UER). The quarterly regression (Panel B) starts from 1947. The sample ends in 1972.
}

\bigskip
\centering

A. Monthly\\
\resizebox*{0.8\textwidth}{!}{\input{tables/table7-ver2.tex}}

\vspace{.3cm}
B. Quarterly \\
\resizebox*{0.8\textwidth}{!}{\input{tables/table7b-ver2.tex}}

\end{table}

\clearpage
\newpage
\begin{table}[!htb]
\caption{EBP Before Historical Crises}\label{tab:ebp-historical-crisis}
{\small
    In this table, we report the average $\hat{\text{EBP}}$ in the 12 months before the start of historical banking panics and NBER recessions defined by \cite{BD:20}. The last row reports the average $\hat{\text{EBP}}$ in all other time periods in 1889-1972. 
}

\centering
\bigskip
\resizebox*{0.5\textwidth}{!}{\input{tables/table8.tex}}
\end{table}

\clearpage
\newpage
\begin{table}[!htb]
\caption{EBP-IG and EBP-HY}\label{tab:ig-hy}
{\small
    This table reports the macroeconomic forecasting regressions using rating-based subsamples of credit spreads. Panel A reports the regression coefficients and standard errors of the forecasting regressions using excess bond premiums averaged separately for investment grade bonds ($\hat{\text{EBP-IG}}$) and Panel B reports the regression using those of high-yield bonds ($\hat{\text{EBP-HY}}$):
    \begin{align*}
        \nabla^h Y_{t+h} &= \alpha + \sum_{i=1}^{p=3} \beta_i \nabla Y_{t-i} + \gamma_1 TS_t + \gamma_2 RFF_t + \gamma_3 GZF_t + \gamma_4 \hat{EBP}_t + \gamma_5 \hat{EBP}\mbox{-}RES_t + \varepsilon_{t+h}, \\
        \nabla^h Y_{t+h} &= \alpha + \sum_{i=1}^{p=4} \beta_i \nabla Y_{t-i} + \gamma_1 TS_t + \gamma_2 RFF_t + \gamma_3 GZF_t + \gamma_4 \hat{EBP}_t + \gamma_5 \hat{EBP}\mbox{-}RES_t + \varepsilon_{t+h},\\
        P(NBER_{t:t+h}) &= \Phi(\alpha + \gamma_1 TS_t + \gamma_2 RFF_t + \gamma_3 GZF_t + \gamma_4 \hat{EBP}_t + \gamma_5. \hat{EBP}\mbox{-}RES_t + \varepsilon_{t+h})
    \end{align*}
    The sample ends on 2022-12-01 due to the availability of the EBP-IG and HY.
}

\bigskip
\centering

A. $\hat{\text{EBP-IG}}$\\
\resizebox*{0.8\textwidth}{!}{\input{tables/table9.tex}}

\vspace{.3cm}
B. $\hat{\text{EBP-HY}}$ \\
\resizebox*{0.8\textwidth}{!}{\input{tables/table9b.tex}}

\end{table}

\clearpage
\newpage
\begin{table}[!htb]
\caption{Topics of Attention (BKMX)}\label{tab:topic-attention-bkmx}
{\small
    This table reports the aggregated weights and explained variance of metatopics for text-based excess
bond premium, $\hat{\text{EBP}}$, and the projection of the fundamental component of the GZ spread on texts,
$\hat{\text{GZF}}$ over the time window 1973-01-01 to 2023-12-01. We aggregate the 180 topics of \cite{bkmx:21} into their 11 metatopics defined and report the aggregated explained variance and coefficients for each
of the metatopics.
}

\bigskip
\centering

A. $\hat{\text{EBP}}$\\
\resizebox*{1\textwidth}{!}{\input{tables/table3_EBP_base_bkmx.tex}}

\vspace{.3cm}
B. $\hat{\text{GZF}}$\\
\resizebox*{1\textwidth}{!}{\input{tables/table3_GZF_base_bkmx.tex}}
\end{table}

\clearpage
\newpage
\begin{table}[!htb]
\caption{Predicting Macroeconomic Fluctuations with Topic Decomposition (BKMX)}\label{tab:topic-decomp-prediction-bkmx}
{\small
This table reports the regression coefficients and standard errors of the three-month-ahead macroeconomic
forecasts in Table~\ref{tab:prediction-macro-fluctuations}, replacing $\hat{\text{EBP}}$ with the excess bond premiums driven by each of the 11 metatopics categorized by \cite{bkmx:21}. The standard errors account for the Newy-West three lags
for EMP-SA, UER, and IPM, one lag for GDP, and 12 lags for Recession.
}

\bigskip
\centering

\resizebox*{1\textwidth}{!}{\input{tables/table4-ver2-bkmx.tex}}

\end{table}

\clearpage
\newpage
\begin{figure}[!htb]
\caption{Summary Statistics of Articles and Lasso}\label{fig:summary-stats}
{\small
    Panel A shows the average number of articles per day in each month. The news data from 1892-01-01 to 1892-06-01 are missing. Panel B shows the positive/negative/non-zero weight topics selected by Lasso regression over time. 
}

\centering
A. Average Number of Articles per Day over a Month\\
\includegraphics[width=\linewidth]{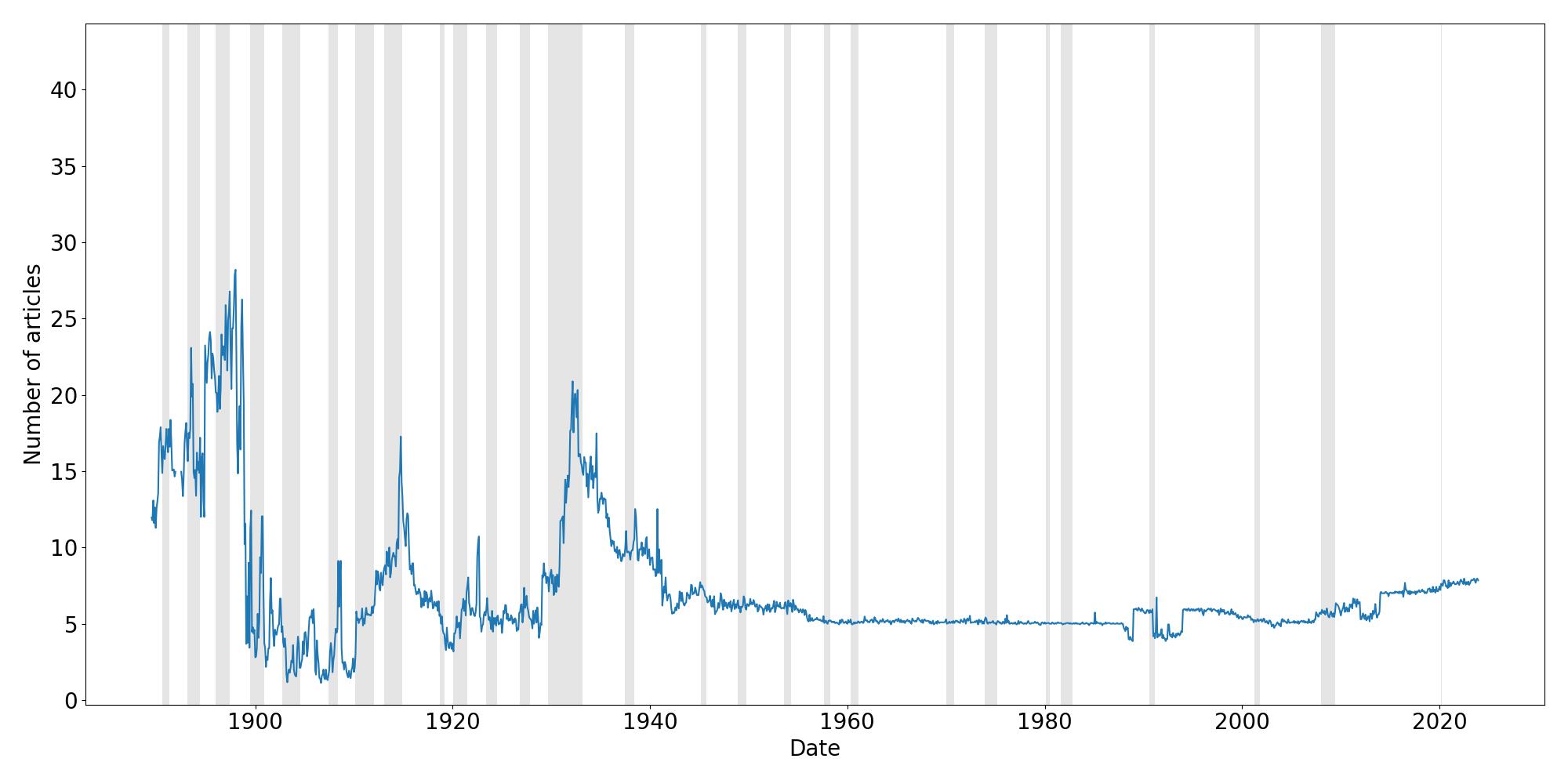}

\vspace{.3cm}
B. Lasso Weights across Time\\
\includegraphics[width=\linewidth]{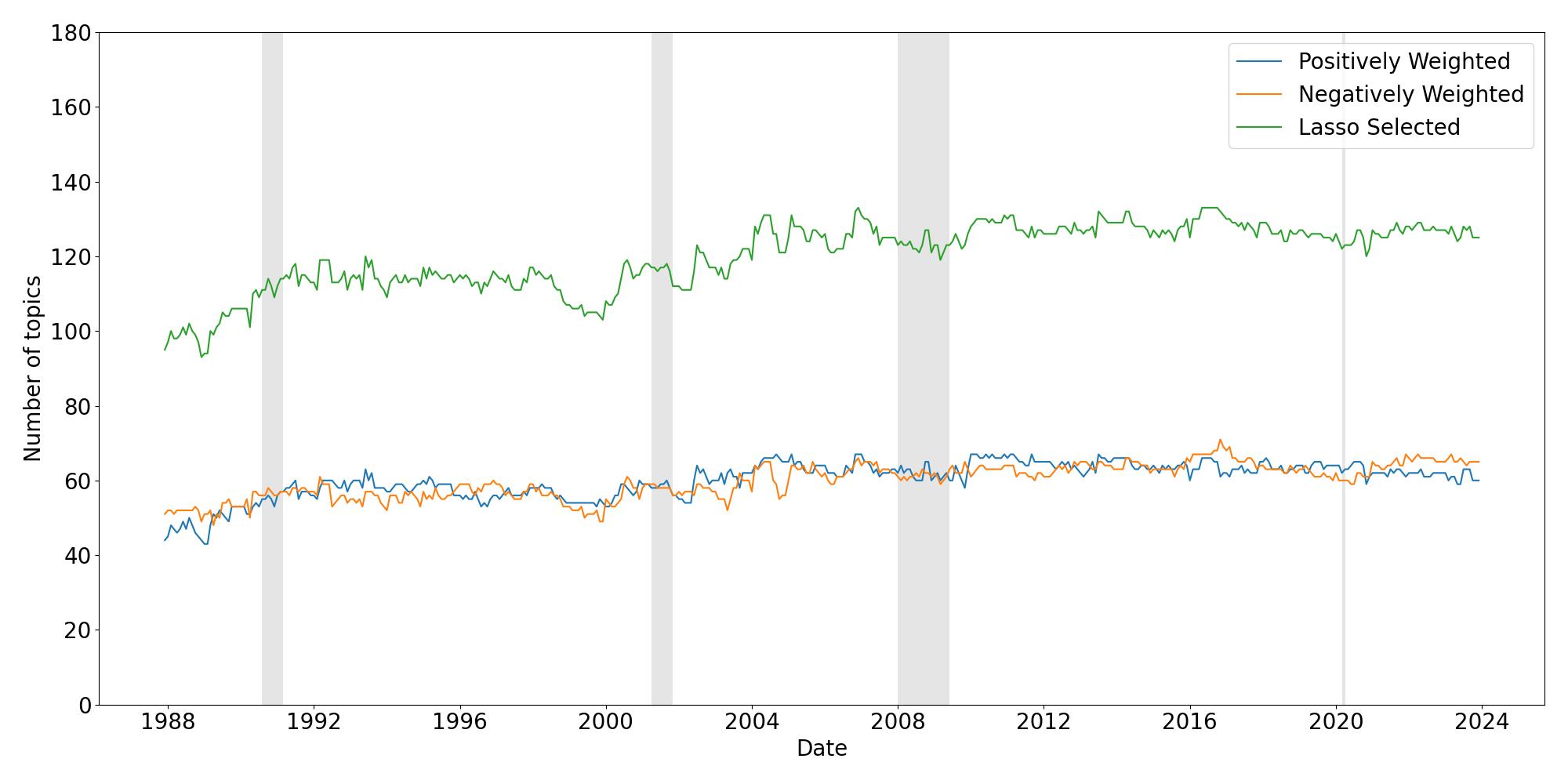}
\end{figure}

\clearpage
\newpage
\begin{figure}[!htb]
\caption{EBP Fitting}\label{fig:ebp-fitting}
{\small These figures plot the excess bond premium (EBP) of \cite{GZ:12}, and their text-based projection created by in-sample forecasts ($\hat{\text{EBP-IS}}$) and out-of-sample forecasts ($\hat{\text{EBP-OOS}}$). Panel A plots the three series over time, while Panel B plots EBP against its text-based projection.  In Panel C, we also plot $\hat{\text{EBP-RES}}$, the difference between the true EBP and text-predicted value.}

\centering
A. Timeseries of $\hat{\text{EBP}}$\\
\includegraphics[width=\linewidth]{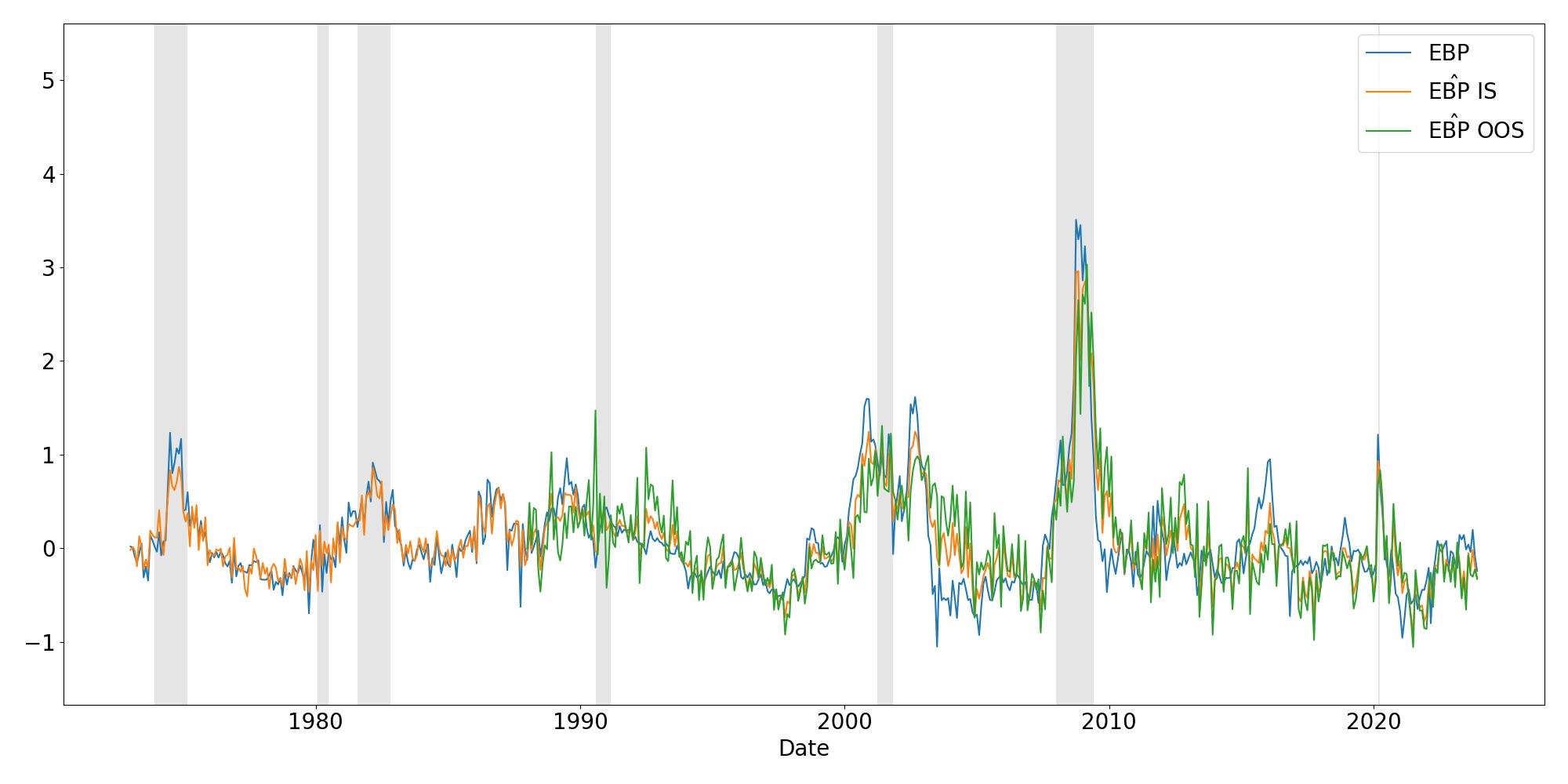}

\vspace{.3cm}
B. Scatter Plots of $\hat{\text{EBP}}$\\
\includegraphics[width=\linewidth]{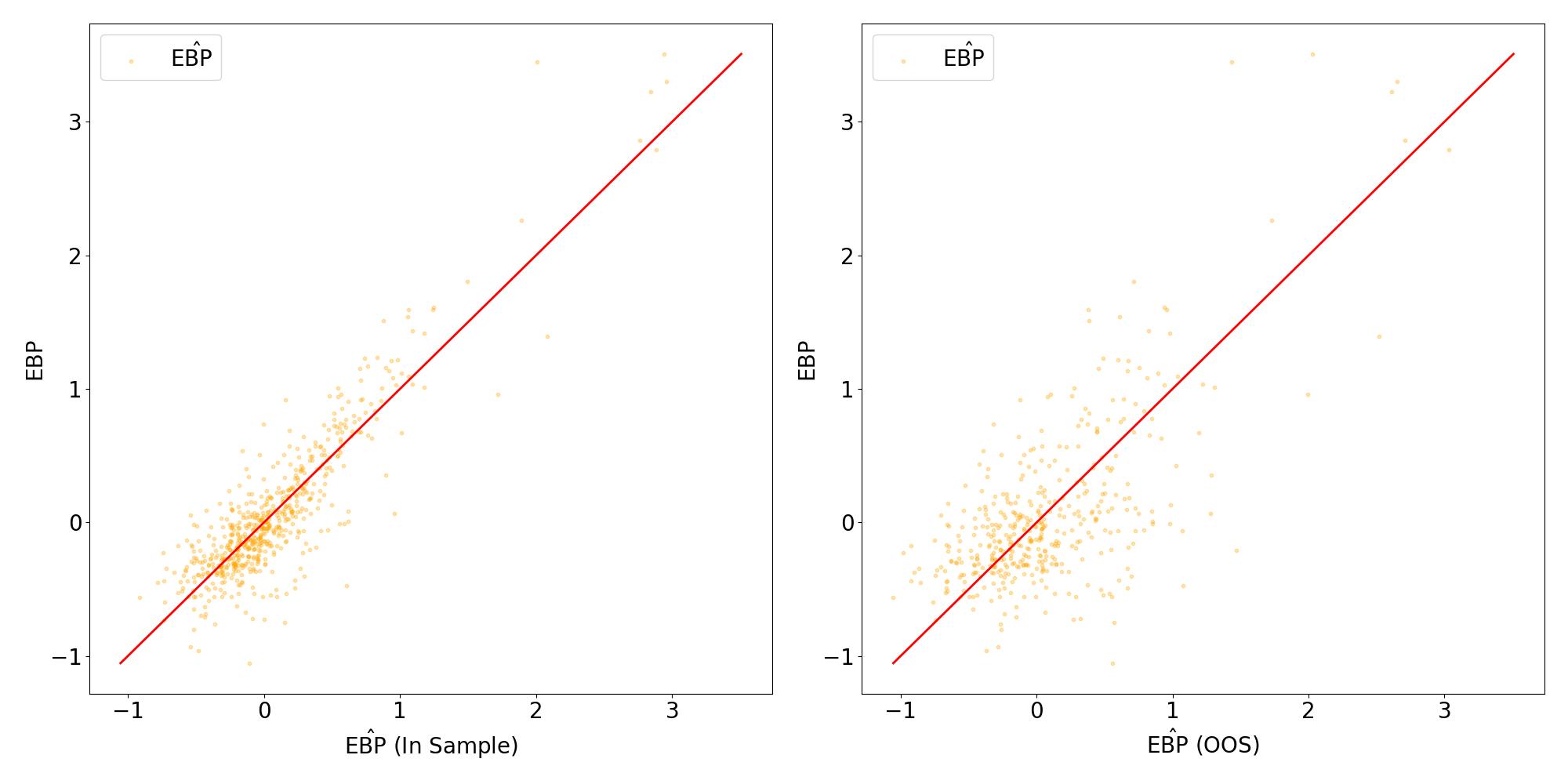}

\end{figure}

\clearpage
\newpage
\begin{figure*}[!htb]
\centering

{
    Figure~\ref{fig:ebp-fitting} continued.
}

C. Timeseries of $\hat{\text{EBP}}$ and $\hat{\text{EBP}}$-RES\\
\includegraphics[width=\linewidth]{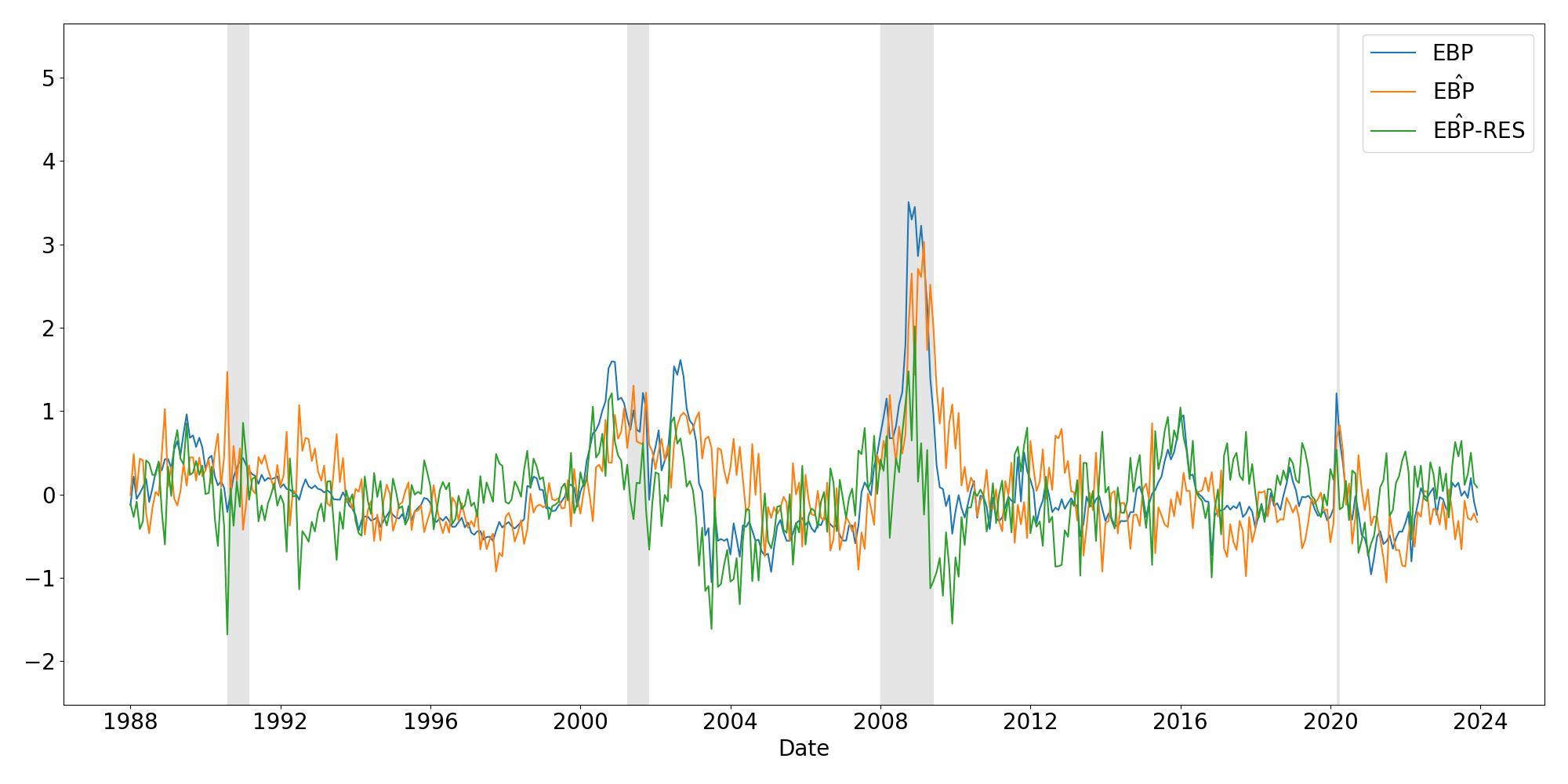}
\end{figure*}

\clearpage
\newpage
\begin{figure}[!htb]
\caption{Recession Probability Estimation}\label{fig:recession-prob}
{\small
    This figure shows the estimated recession probability with the following probit regressions:
    \begin{align*}
    P(NBER_{t:t+h}) &= \Phi(\alpha + \gamma_1 TS_t + \gamma_2 RFF_t + \gamma_3 GZF_t + \gamma_4 EBP_t + \varepsilon_{t+h})\\
    P(NBER_{t:t+h}) &= \Phi(\alpha + \gamma_1 TS_t + \gamma_2 RFF_t + \gamma_3 GZF_t + \gamma_4 \hat{EBP}_t + \varepsilon_{t+h})
    \end{align*}
    where $h=3$.  The line EBP corresponds to the forecast using EBP, while the line $\hat{\text{EBP}}$ plots the forecast using $\hat{\text{EBP}}$. 
}

\includegraphics[width=\linewidth]{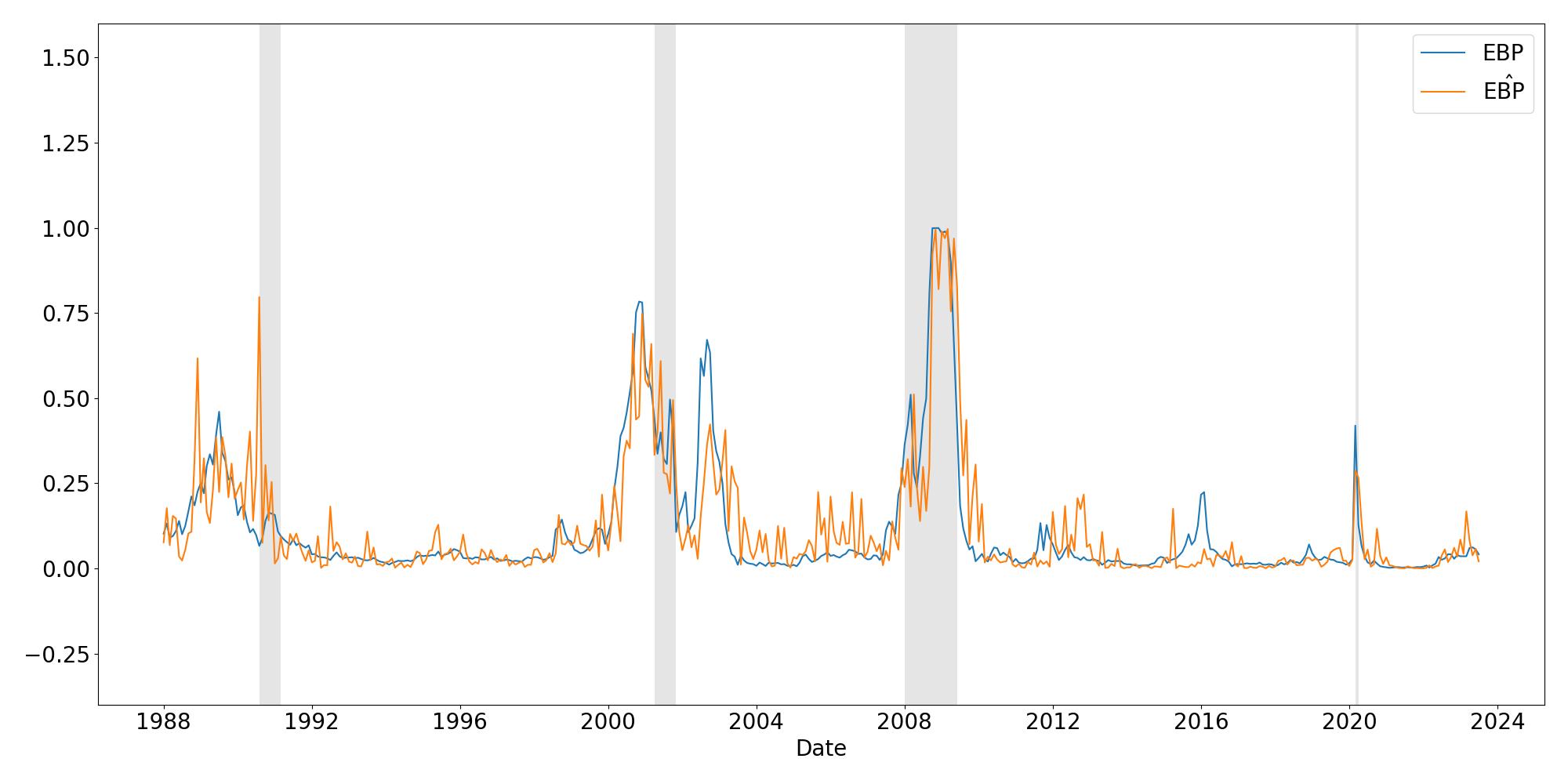}
\end{figure}

\clearpage
\newpage
\begin{figure}[!htb]
\caption{Attention Weights on Metatopic across Time}\label{fig:metatopic-time}
{\small This figure plots estimated time variation in attention weights on each metatopic defined by GPT-o1 on an expanding basis.  Panel A plots the weights on each topic that explains excess bond premium (EBP) and Panel B plots the weights for the fundamental-based credit spreads (GZF).}

\centering
A. $\hat{\text{EBP}}$\\
\includegraphics[width=0.6\linewidth]{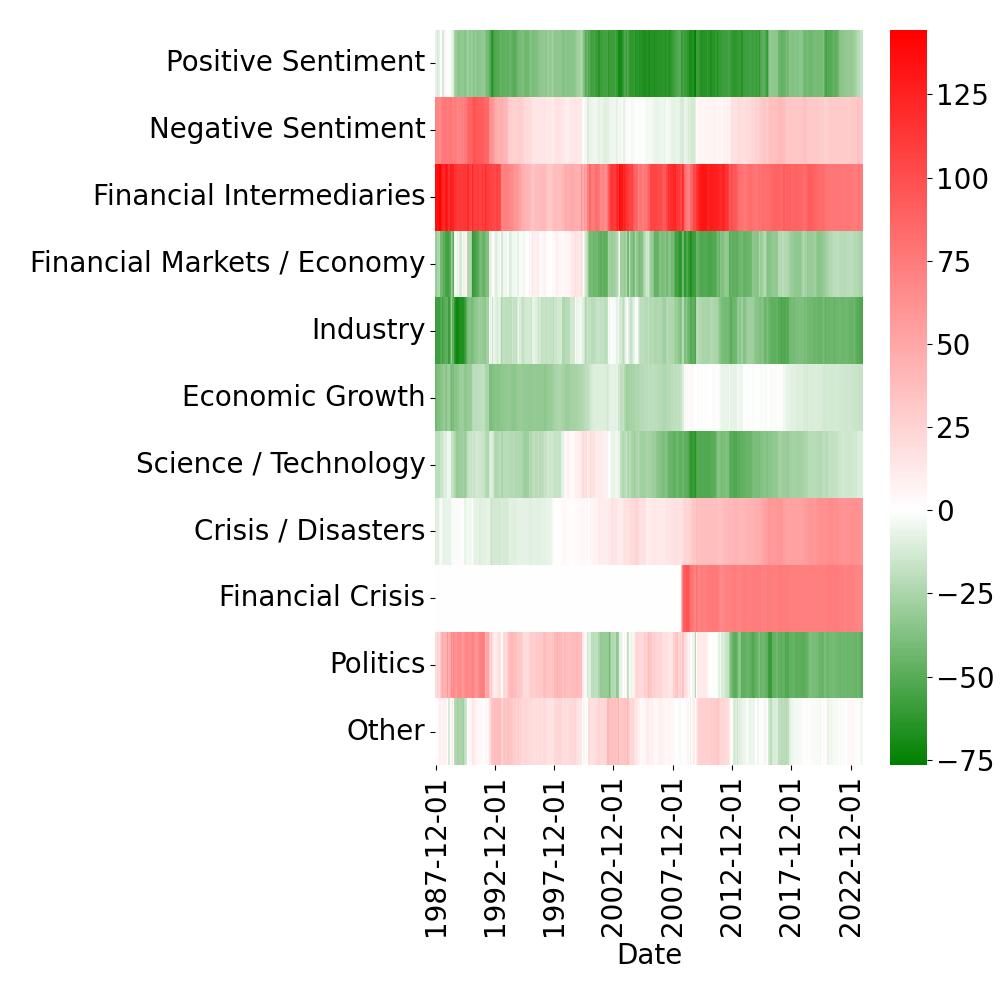}

\vspace{.3cm}
B. $\hat{\text{GZF}}$\\
\includegraphics[width=0.6\linewidth]{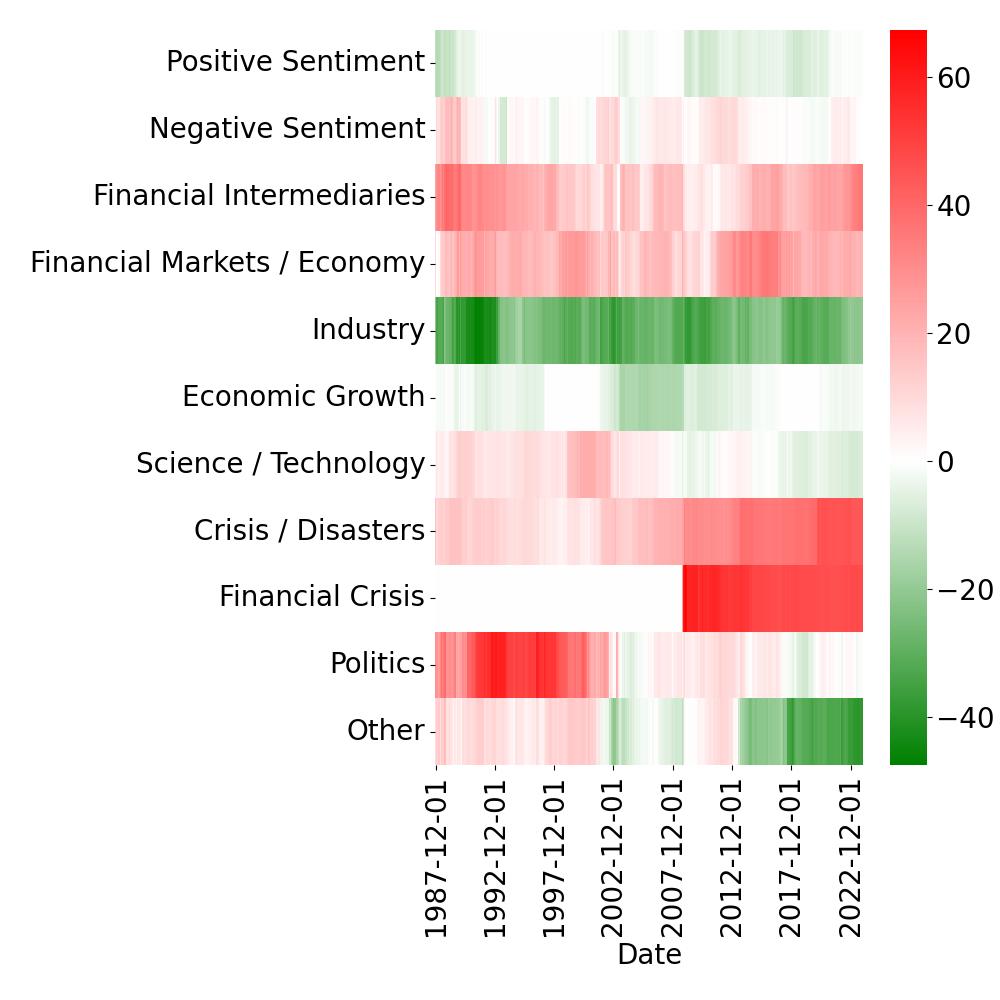}
\end{figure}

\clearpage
\newpage
\begin{figure}[!htb]
\caption{Topic Decomposition}\label{fig:topic-decomposition}
{\small This figure plots the excess bond premiums driven by metatopics.  For every metatopic $M$, we take all topics $k \in M$ and calculate the component of EBP associated specifically with that metatopic, $\hat{\text{EBP[M]}}_t \equiv \sum_k w_{k,t-1}\theta_{k,t}$, where $w_{k,t-1}$ and $\theta_{k,t}$ equal the (stale) model weight and news attention for topic k, respectively.}

\centering
\includegraphics[width=\linewidth]{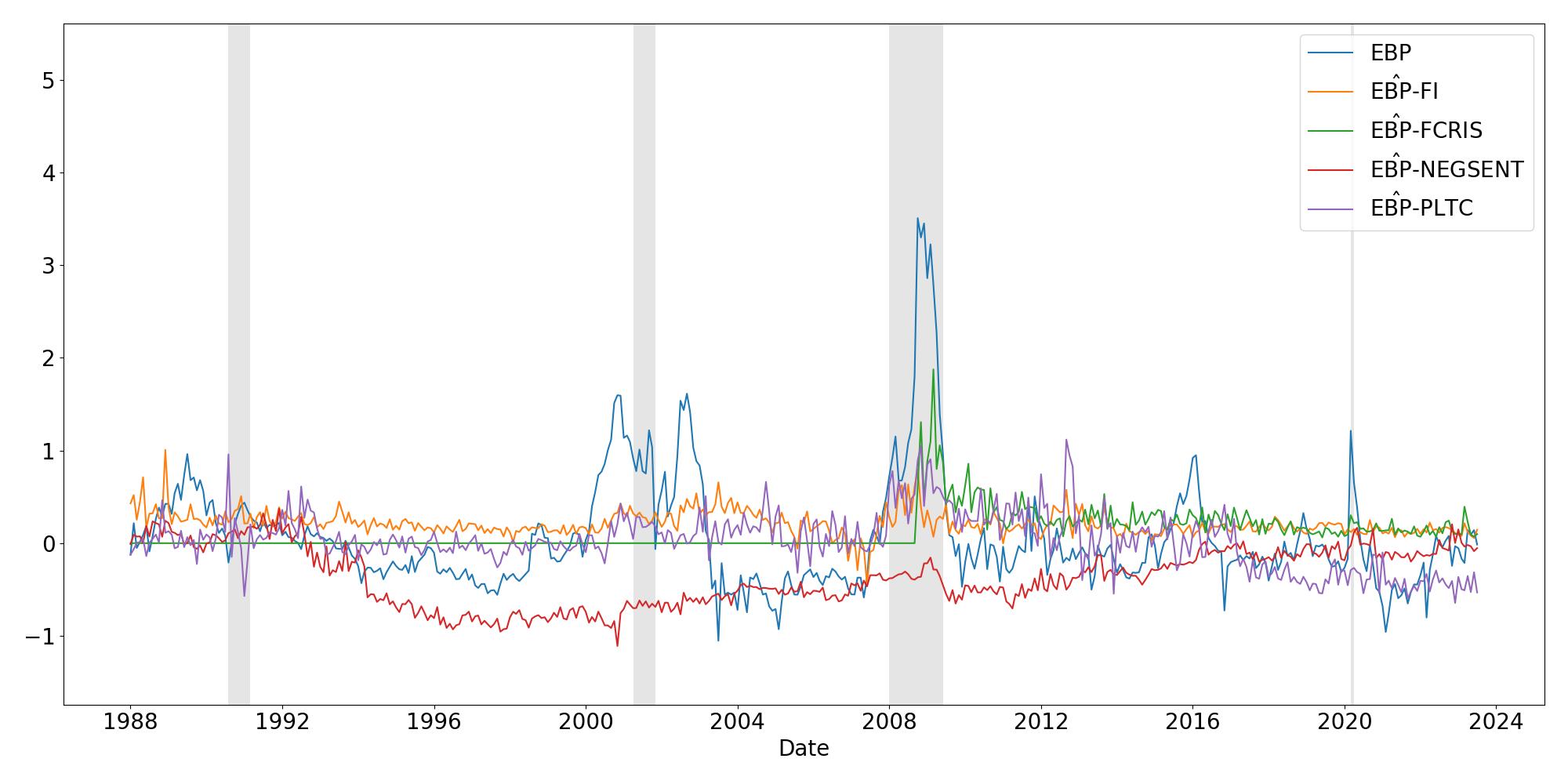}
\end{figure}

\clearpage
\newpage
\begin{figure}[!htb]
\caption{Sentiment}\label{fig:sentiment}
{\small This figure plots the sentiment-based excess bond premiums. $SENT^{LM}$ is the sentiment measure based on \cite{LM:11} dictionaries using all words in a month, $SENT^A$ is a topic weighted sentiment score, and $\hat{\hat{EBP}}$ is the fitted value of $EBP$ using $SENT^{LM}$.}

\centering
A. $\text{SENT}^{LM}$ vs. $\text{SENT}^A$\\
\includegraphics[width=\linewidth]{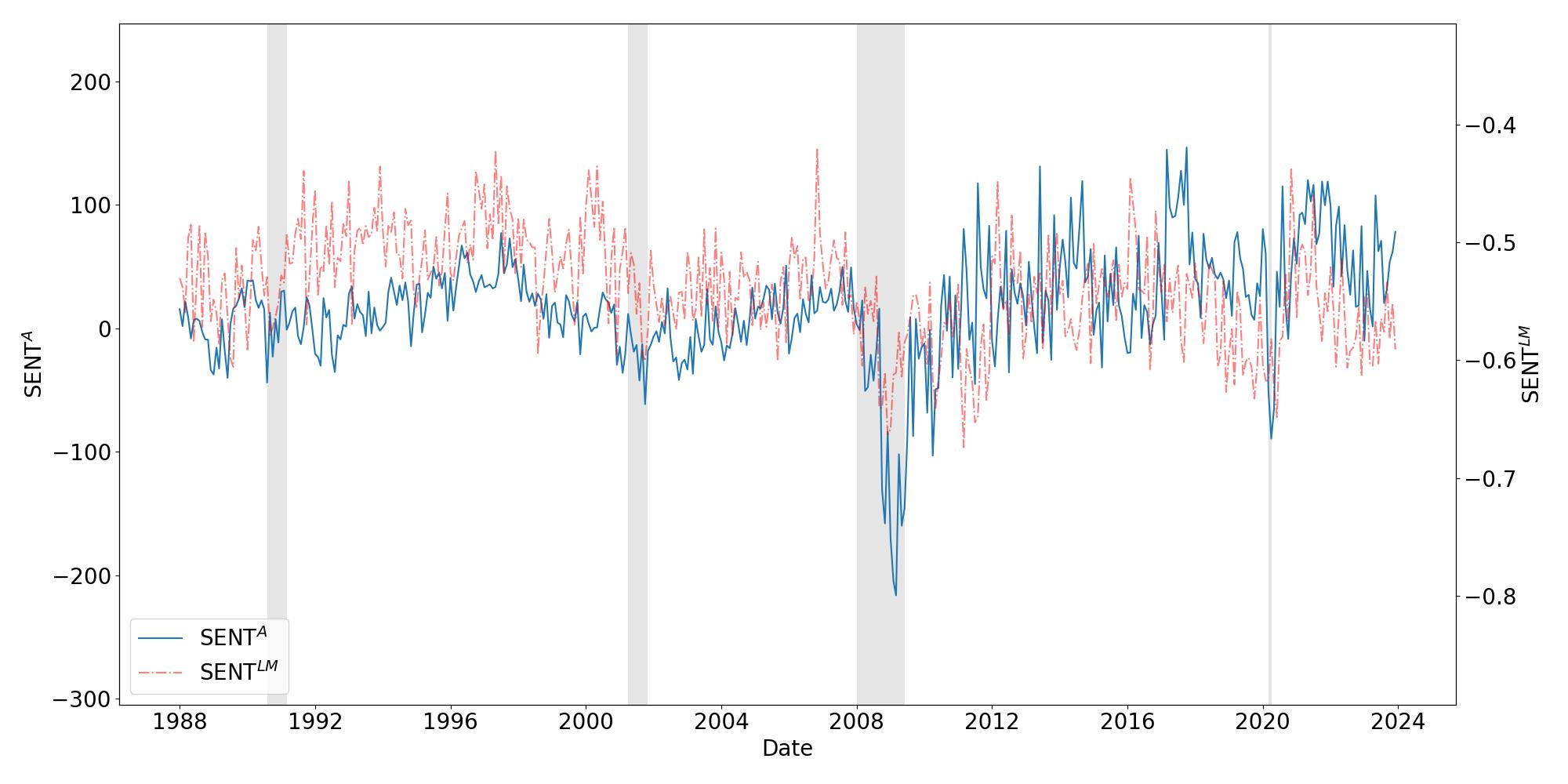}

\vspace{.3cm}
B. $\text{EBP}$ vs. $\hat{\hat{\text{EBP}}}$\\
\includegraphics[width=\linewidth]{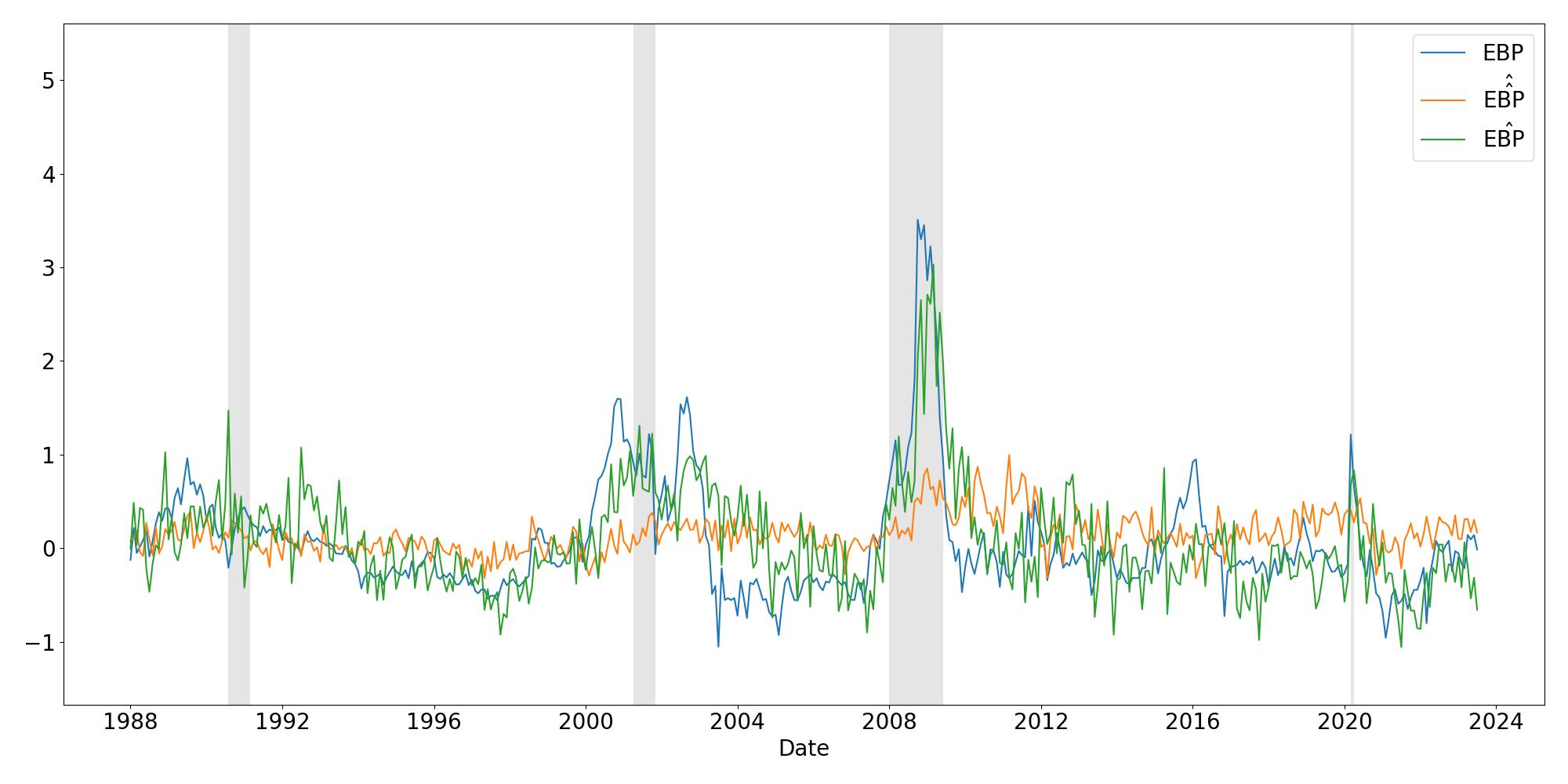}
\end{figure}

\clearpage
\newpage
\begin{figure}[!htb]
\caption{Backward Projection}\label{fig:backward-projection}
{\small This figure plots out-of-sample prediction ($\hat{\text{EBP}}$ OOS) from July 1889 to December 1972
using the period of January 1973 to December 2023 as the in-sample training period.  The figure also plots the in-sample fitted values ($\hat{\text{EBP}}$ IS) over the training period.
}

\centering
\includegraphics[width=\linewidth]{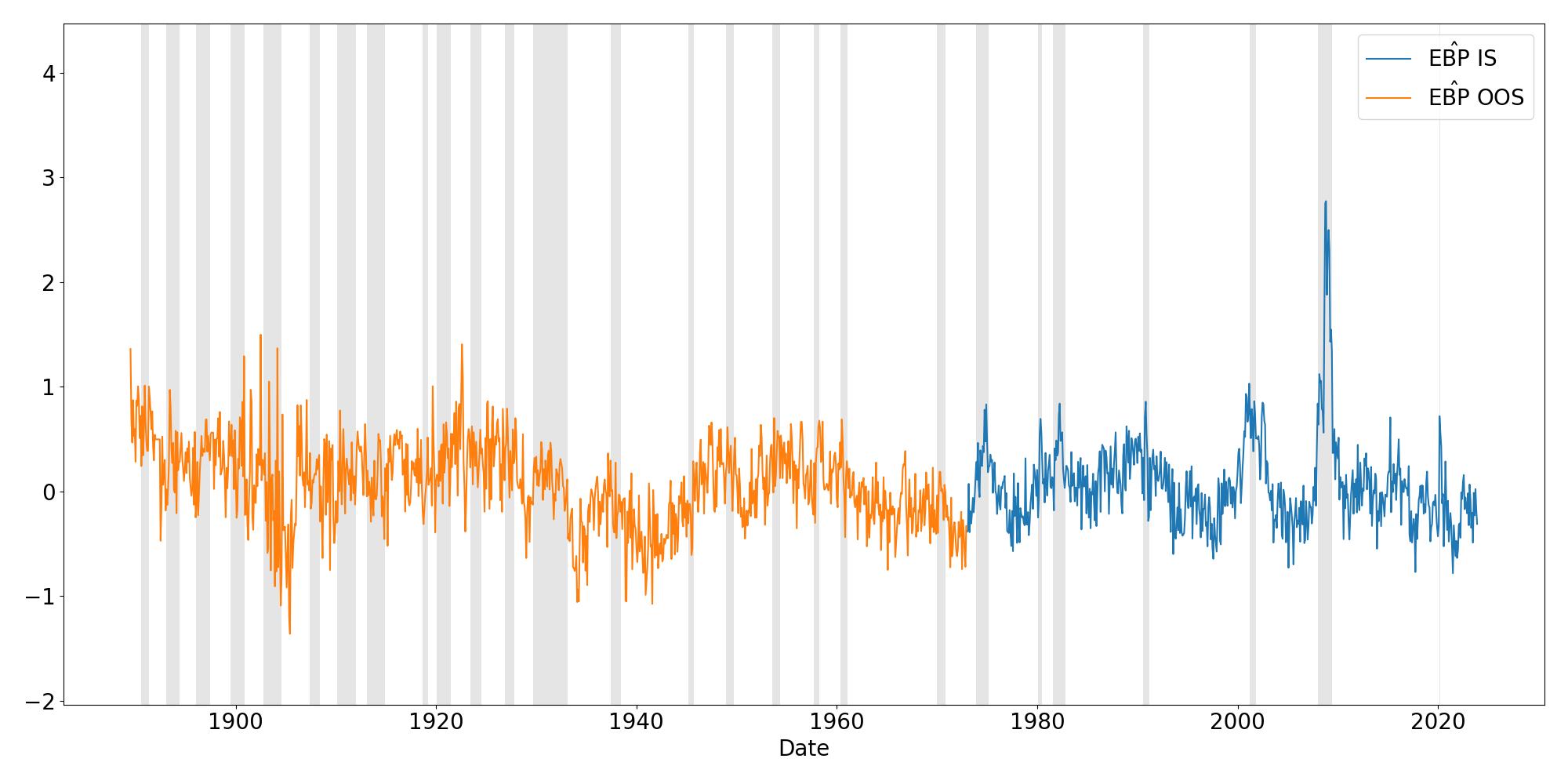}

\end{figure}

\clearpage
\newpage
\begin{figure}[!htb]
\caption{Attention Weights on Metatopic across Time (BKMX)}\label{fig:metatopic-time-bkmx}
{\small This figure plots estimated time variation in attention weights on each metatopic defined by \cite{bkmx:21} on a rolling basis. Panel A plots the
weights on each topic that explains excess bond premium (EBP) and Panel B plots the weights for
the fundamental-based credit spreads (GZF).}

\centering
A. $\hat{\text{EBP}}$\\
\includegraphics[width=0.6\linewidth]{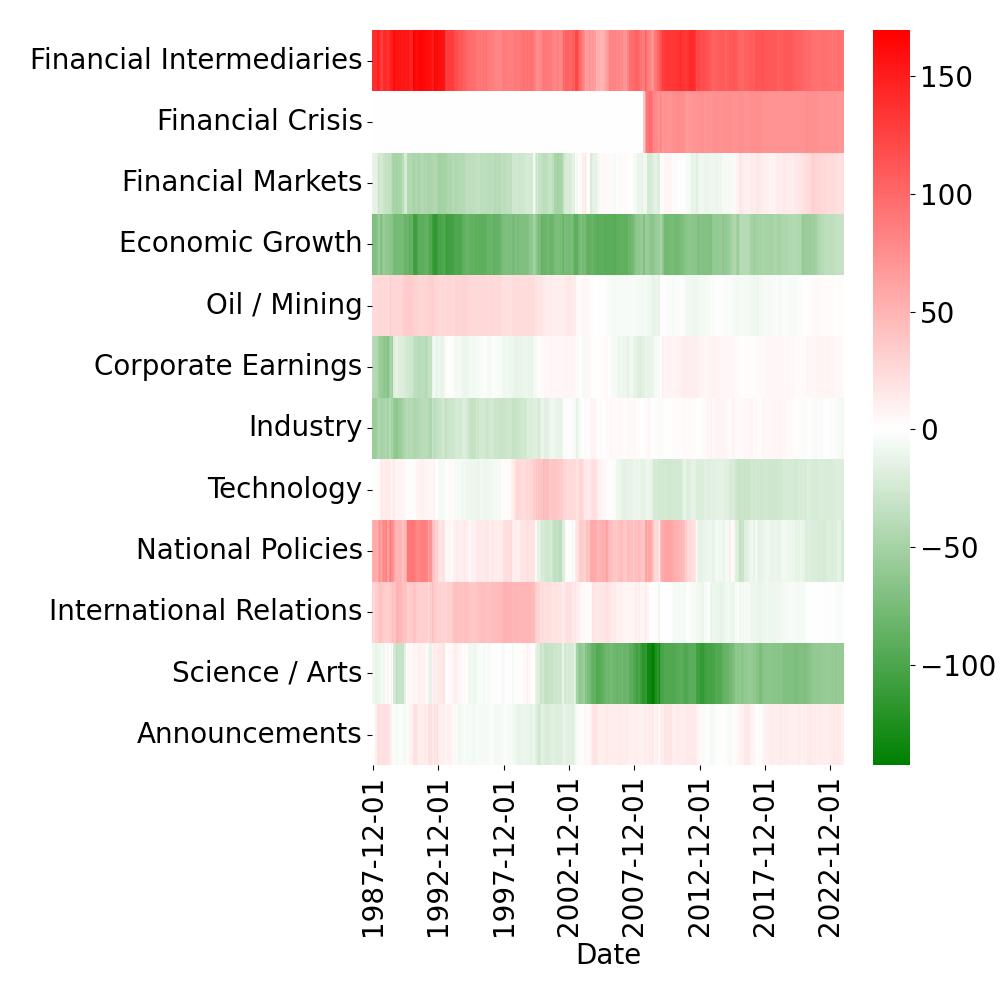}

\vspace{.3cm}
B. $\hat{\text{GZF}}$\\
\includegraphics[width=0.6\linewidth]{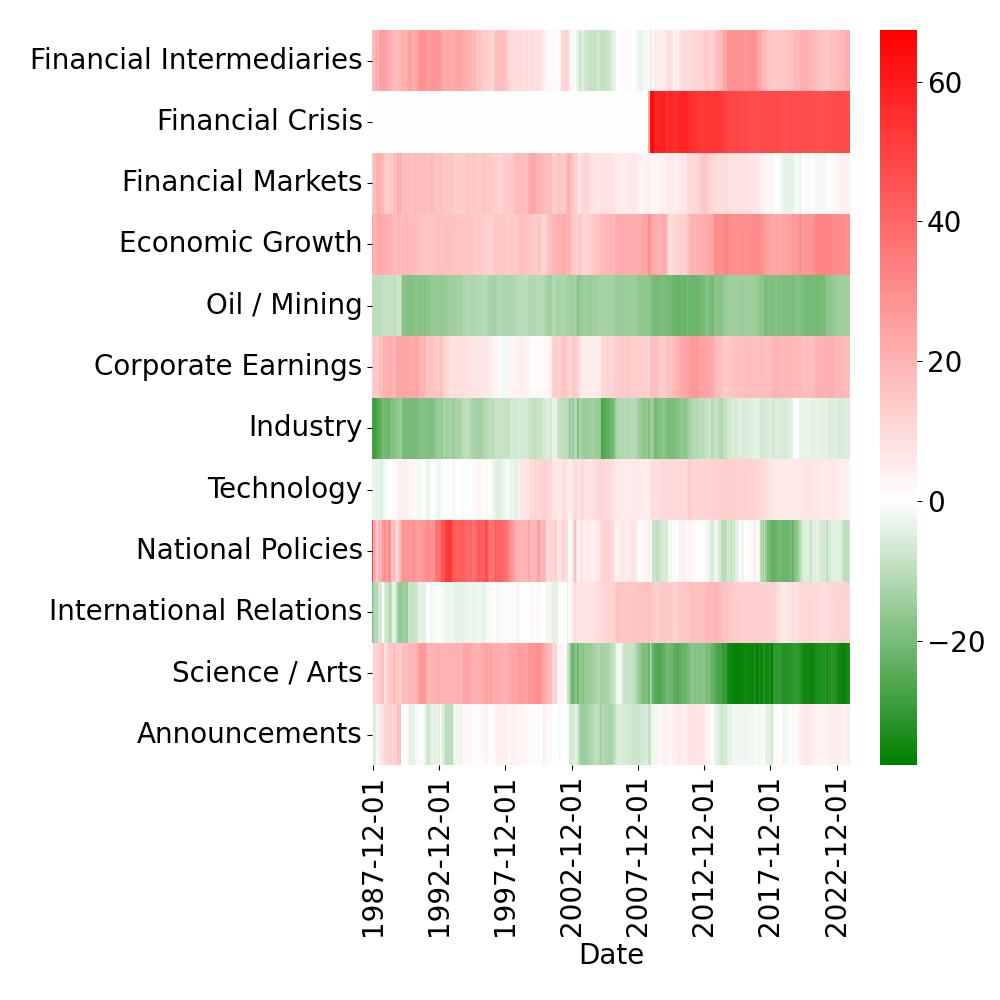}
\end{figure}

%% file: tables/table1.tex
\begin{tabular}{lcccc}
\hline\hline
 & IS, Training (1973m1-1987m12) & IS (1973m1+) & IS, Post-training (1988m1+) & OOS (1988m1+) \\
 & (1) & (2) & (3) & (4) \\
\cmidrule(lr){2-4} \cmidrule(lr){5-5}
$\hat{\text{EBP}}$ & 1.1185*** & 1.0094*** & 0.9953*** & 0.7275*** \\
 & (0.05) & (0.05) & (0.06) & (0.13) \\
Intercept & -0.0086 & -0.0025 & -0.0025 & 0.0055 \\
 & (0.01) & (0.02) & (0.02) & (0.04) \\
$R^2$ & 0.861 & 0.786 & 0.776 & 0.445 \\
$T$ & 180 & 612 & 432 & 432 \\
RMSE & 0.139 & 0.253 & 0.287 & 0.477 \\
MAE & 0.110 & 0.182 & 0.212 & 0.360 \\
\hline
\end{tabular}

%% file: tables/table2-ver2.tex
\begin{tabular}{lccccc}
\hline\hline
 & EMP (SA) & UER & IPM & GDP & Recession \\
 & (1) & (2) & (3) & (4) & (5) \\
\cmidrule(lr){2-4} \cmidrule(lr){5-5} \cmidrule(lr){6-6}
TS & 0.3032 & -0.2148 & 1.5547** & 0.1713 & -0.130 \\
 & (0.49) & (0.37) & (0.67) & (0.41) & (0.194) \\
RFF & 0.0590 & 0.0946 & 0.5942* & 0.1322 & 0.087 \\
 & (0.23) & (0.17) & (0.33) & (0.20) & (0.119) \\
GZF & -1.0055** & 0.4685* & -0.4662 & -0.6819 & 0.254 \\
 & (0.40) & (0.28) & (1.06) & (0.60) & (0.510) \\
$\hat{\text{EBP}}$ & -2.7400*** & 1.8126*** & -6.8227*** & -2.4496*** & 1.669*** \\
 & (0.63) & (0.48) & (1.56) & (0.62) & (0.276) \\
$\hat{\text{EBP}}$-RES & -1.7091*** & 1.2820*** & -7.2367*** & -2.6224*** & 0.797** \\
 & (0.37) & (0.28) & (1.58) & (0.60) & (0.324) \\
$R^2$ & 0.162 & 0.145 & 0.262 & 0.183 & 0.372 \\
$T$ & 427 & 427 & 427 & 142 & 427 \\
SHAP(GZF) & 14.67 & 8.98 & 2.08 & 9.47 & 7.22 \\
SHAP($\hat{\text{EBP}}$) & 36.57 & 31.78 & 27.87 & 32.76 & 52.92 \\
SHAP($\hat{\text{EBP}}$-RES) & 20.24 & 19.95 & 26.24 & 30.70 & 19.67 \\
\hline
\end{tabular}

%% file: tables/table2b.tex
\begin{tabular}{lccccc}
\hline\hline
 & EMP (SA) & UER & IPM & GDP & Recession \\
 & (1) & (2) & (3) & (4) & (5) \\
\cmidrule(lr){2-4} \cmidrule(lr){5-5} \cmidrule(lr){6-6}
TS & 0.1488 & -0.1352 & 1.6206** & 0.1921 & 0.016 \\
 & (0.47) & (0.36) & (0.66) & (0.41) & (0.208) \\
RFF & 0.0018 & 0.1230 & 0.6193* & 0.1386 & 0.138 \\
 & (0.22) & (0.17) & (0.32) & (0.21) & (0.125) \\
GZF & -1.3432*** & 0.6446** & -0.3250 & -0.6286 & 0.519 \\
 & (0.39) & (0.26) & (0.98) & (0.56) & (0.467) \\
EBP & -2.1920*** & 1.5331*** & -7.0580*** & -2.5435*** & 1.160*** \\
 & (0.47) & (0.35) & (1.47) & (0.55) & (0.268) \\
$R^2$ & 0.157 & 0.143 & 0.264 & 0.189 & 0.337 \\
$T$ & 427 & 427 & 427 & 142 & 427 \\
SHAP(GZF) & 29.13 & 15.93 & 1.92 & 12.19 & 18.24 \\
SHAP(EBP) & 43.63 & 34.77 & 38.23 & 48.53 & 56.28 \\
\hline
\end{tabular}

%% file: tables/table3_EBP_base_gpt.tex
\begin{tabular}{llcc}
\toprule
 &  & Explained Variance & Raw Weight \\
\cmidrule(lr){2-2} \cmidrule(lr){3-4}
\multirow[t]{4}{*}{Top Positive} & Crisis / Disasters & 9.97 & 60.2013 \\
 & Financial Crisis & 6.22 & 69.6270 \\
 & Negative Sentiment & 2.82 & 31.8725 \\
 & Financial Intermediaries & 2.61 & 73.6411 \\
\multirow[t]{7}{*}{Top Negative} & Politics & 28.16 & -48.5888 \\
 & Other & 22.94 & -4.7921 \\
 & Industry & 11.46 & -53.5956 \\
 & Financial Markets / Economy & 11.31 & -26.2482 \\
 & Positive Sentiment & 3.23 & -17.2497 \\
 & Science / Technology & 0.79 & -10.2909 \\
 & Economic Growth & 0.49 & -17.1344 \\
\bottomrule
\end{tabular}

%% file: tables/table3_GZF_base_gpt.tex
\begin{tabular}{llcc}
\toprule
 &  & Explained Variance & Raw Weight \\
\cmidrule(lr){2-2} \cmidrule(lr){3-4}
\multirow[t]{5}{*}{Top Positive} & Financial Markets / Economy & 30.47 & 19.8485 \\
 & Crisis / Disasters & 7.14 & 44.4904 \\
 & Negative Sentiment & 6.90 & 0.0881 \\
 & Financial Intermediaries & 4.09 & 36.0930 \\
 & Financial Crisis & 4.09 & 47.2805 \\
\multirow[t]{6}{*}{Top Negative} & Other & 19.11 & -38.9193 \\
 & Politics & 13.34 & -0.5528 \\
 & Industry & 13.02 & -20.7913 \\
 & Science / Technology & 1.83 & -6.9215 \\
 & Positive Sentiment & 0.02 & -0.6818 \\
 & Economic Growth & 0.01 & -2.0806 \\
\bottomrule
\end{tabular}

%% file: tables/representative_articles.tex
\begin{tabular}{p{0.2\linewidth}p{0.15\linewidth}clc}
\toprule
Topic & Month with  & $\hat{\text{EBP}}$-M & Article Title & Topic\\
 & Top $\hat{\text{EBP}}$-M &  &  & Attention\\
\midrule
\multirow{2}{12pt}{Financial Intermediaries} & 1988-12 & 1.007 & It Could Become A Novel: 'Campfire Of the Vanities' (1988-12-22) & 0.4194\\
 & & & Drexel's Chief Faces Decision That Offers Little Solace for Firm (1988-12-19) & 0.4167\\
 & & & A Modern Merlin Helps Corporations Work Their Magic (1988-12-29) & 0.4054\\ \\
Financial Crisis & 2009-03 & 1.8761 & Geithner Banks on Private Cash (2009-03-23) & 0.2424\\
 & & & U.S. to Toughen Finance Rules (2009-03-16) & 0.1569\\
 & & & Treasury Maps New Era of Regulation (2009-03-27) & 0.1270\\ \\
Politics & 2012-09 & 1.1181 & Romney Offers New Tax Details (2012-09-22) & 0.7423\\
& & & Israel Blasts U.S. Over Iran (2012-09-12) & 0.7217\\
& & & Netanyahu Demands 'Red Line' on Iran (2012-09-28) & 0.6606\\ \\
Crisis /Disasters & 2020-05 & 0.6646 & Coronavirus Hijacks the Body From Head to Toe (2020-05-08) & 0.3034\\
& & & With Death Rates on the Rise, Precise Virus Toll Still Murky (2020-05-16) & 0.2727\\
& & & China Stalls Global Hunt for Origins of Virus in Wuhan (2020-05-13) & 0.2645\\
\bottomrule
\end{tabular}

%% file: tables/table4-ver2-gpt.tex
\begin{tabular}{lccccc}
\hline\hline
 & EMP (SA) & UER & IPM & GDP & Recession \\
 & (1) & (2) & (3) & (4) & (5) \\
\cmidrule(lr){2-4} \cmidrule(lr){5-5} \cmidrule(lr){6-6}
TS & 0.3974 & -0.3720 & 2.3303*** & 0.9050* & -0.146 \\
 & (0.54) & (0.40) & (0.81) & (0.53) & (0.220) \\
RFF & 0.0593 & 0.0040 & 0.8977** & 0.5549 & 0.185 \\
 & (0.28) & (0.20) & (0.45) & (0.34) & (0.177) \\
GZF & -2.4859** & 1.1512 & -2.5398 & -0.1760 & 1.149** \\
 & (1.11) & (0.81) & (1.60) & (1.21) & (0.487) \\
$\hat{\text{EBP}}$-CRISDIS & 0.1675 & 0.2105 & 4.6125 & 3.1557 & 0.484 \\
 & (3.96) & (3.00) & (6.31) & (3.99) & (1.357) \\
$\hat{\text{EBP}}$-GROWTH & -1.6765 & -0.2732 & 13.2622 & -7.0579 & 2.231 \\
 & (6.52) & (4.55) & (13.29) & (10.58) & (5.488) \\
$\hat{\text{EBP}}$-FCRIS & -1.9954** & 1.7195** & -4.3125 & -2.3554 & 0.555 \\
 & (0.98) & (0.82) & (3.26) & (2.38) & (0.896) \\
$\hat{\text{EBP}}$-FI & -0.3999 & 0.5927 & -14.7330*** & -4.4834* & 1.588** \\
 & (1.42) & (1.08) & (3.57) & (2.43) & (0.763) \\
$\hat{\text{EBP}}$-MKTECO & -2.8832* & 2.5244** & -9.2459** & -2.2626 & 2.079** \\
 & (1.47) & (1.13) & (3.87) & (2.01) & (0.967) \\
$\hat{\text{EBP}}$-IND & 2.2580 & -1.0080 & -4.0083 & 1.4266 & -0.868 \\
 & (2.91) & (2.16) & (3.35) & (2.56) & (0.996) \\
$\hat{\text{EBP}}$-NEGSENT & -5.0957* & 3.2000 & -9.6847*** & -1.0573 & 3.531*** \\
 & (2.77) & (2.04) & (3.34) & (1.92) & (0.835) \\
$\hat{\text{EBP}}$-OTHER & -4.0083 & 3.0294 & -1.2688 & -1.4673 & 1.158 \\
 & (2.88) & (2.11) & (2.87) & (1.60) & (0.856) \\
$\hat{\text{EBP}}$-PLTC & -1.6257* & 1.2926** & -6.9148*** & -4.0261*** & 1.963*** \\
 & (0.92) & (0.66) & (1.86) & (1.11) & (0.550) \\
$\hat{\text{EBP}}$-POSSENT & -4.2328 & 3.4232* & -2.4549 & -9.6099** & -3.495 \\
 & (2.57) & (1.90) & (6.14) & (4.68) & (2.426) \\
$\hat{\text{EBP}}$-SCITECH & -1.3681 & 0.5312 & -5.5853*** & -0.2295 & 1.946*** \\
 & (0.85) & (0.62) & (1.78) & (1.15) & (0.653) \\
$\hat{\text{EBP}}$-RES & -1.2915*** & 1.0827*** & -7.2633*** & -2.5867*** & 0.707** \\
 & (0.32) & (0.23) & (1.35) & (0.51) & (0.325) \\
$R^2$ & 0.149 & 0.133 & 0.329 & 0.188 & 0.417 \\
$T$ & 427 & 427 & 427 & 142 & 427 \\
\hline
\end{tabular}

%% file: tables/table5.tex
\begin{tabular}{lccccccccc}
\hline\hline
  & \multicolumn{3}{c}{$\text{SENT}^{LM}$} & \multicolumn{3}{c}{$\text{SENT}^A$} & \multicolumn{3}{c}{$\hat{\hat{\text{EBP}}}$} \\
 & EMP (SA) & UER & IPM & EMP (SA) & UER & IPM & EMP (SA) & UER & IPM \\
 & (1) & (2) & (3) & (4) & (5) & (6) & (7) & (8) & (9) \\
\cmidrule(lr){2-4} \cmidrule(lr){5-7} \cmidrule(lr){8-10}
TS & -0.1186 & 0.0494 & 0.8019 & 0.4905 & -0.3478 & 1.6462*** & 0.1462 & -0.1340 & 1.6283** \\
 & (0.42) & (0.31) & (0.65) & (0.50) & (0.37) & (0.62) & (0.47) & (0.35) & (0.66) \\
RFF & -0.0456 & 0.1847* & 0.0774 & 0.2392 & 0.0025 & 0.5282** & -0.0593 & 0.1635 & 0.5614* \\
 & (0.15) & (0.11) & (0.22) & (0.20) & (0.15) & (0.26) & (0.19) & (0.15) & (0.30) \\
$\text{SENT}^{LM}$ & 23.1076** & -15.4647** & 43.7778*** &  &  &  &  &  &  \\
 & (9.15) & (6.60) & (15.61) &  &  &  &  &  &  \\
$\text{SENT}^A$ &  &  &  & 0.0325*** & -0.0209*** & 0.0383** &  &  &  \\
 &  &  &  & (0.01) & (0.01) & (0.02) &  &  &  \\
GZF &  &  &  &  &  &  & -1.1886*** & 0.5435* & -0.1477 \\
 &  &  &  &  &  &  & (0.43) & (0.29) & (1.04) \\
$\hat{\hat{\text{EBP}}}$ &  &  &  &  &  &  & -4.7480*** & 3.1640** & -9.7519*** \\
 &  &  &  &  &  &  & (1.73) & (1.24) & (3.00) \\
$\hat{\hat{\text{EBP}}}$-RES &  &  &  &  &  &  & -1.9770*** & 1.3971*** & -6.8723*** \\
 &  &  &  &  &  &  & (0.42) & (0.32) & (1.47) \\
$R^2$ & 0.062 & 0.073 & 0.069 & 0.110 & 0.108 & 0.054 & 0.168 & 0.151 & 0.266 \\
$T$ & 427 & 427 & 427 & 427 & 427 & 427 & 427 & 427 & 427 \\
SHAP($\text{SENT}^{LM}$) & 60.39 & 38.46 & 52.69 &  &  &  &  &  &  \\
SHAP($\text{SENT}^A$) &  &  &  & 41.05 & 44.39 & 25.99 &  &  &  \\
SHAP(GZF) &  &  &  &  &  &  & 18.78 & 10.39 & 0.74 \\
SHAP($\hat{\hat{\text{EBP}}}$) &  &  &  &  &  &  & 25.89 & 20.88 & 16.77 \\
SHAP($\hat{\hat{\text{EBP}}}$-RES) &  &  &  &  &  &  & 29.77 & 25.45 & 32.64 \\
\hline
\end{tabular}

%% file: tables/table6.tex
\begin{tabular}{lccccccccc}
\hline\hline
  & \multicolumn{3}{c}{$\text{SENT}^{LM}$} & \multicolumn{3}{c}{$\text{SENT}^A$} & \multicolumn{3}{c}{$\hat{\hat{\text{EBP}}}$} \\
 & EMP (SA) & UER & IPM & EMP (SA) & UER & IPM & EMP (SA) & UER & IPM \\
 & (1) & (2) & (3) & (4) & (5) & (6) & (7) & (8) & (9) \\
\cmidrule(lr){2-4} \cmidrule(lr){5-7} \cmidrule(lr){8-10}
TS & 0.2030 & -0.1466 & 1.3905** & 0.3461 & -0.2512 & 1.4699** & 0.2867 & -0.2042 & 1.5469** \\
 & (0.43) & (0.33) & (0.61) & (0.50) & (0.38) & (0.68) & (0.48) & (0.36) & (0.66) \\
RFF & -0.0218 & 0.1514 & 0.4552 & 0.1045 & 0.0572 & 0.5087 & -0.0043 & 0.1368 & 0.5289* \\
 & (0.18) & (0.14) & (0.28) & (0.25) & (0.18) & (0.34) & (0.20) & (0.15) & (0.30) \\
GZF & -0.9376** & 0.4212 & -0.3022 & -0.8472* & 0.3347 & -0.7619 & -0.8889** & 0.3926 & -0.3181 \\
 & (0.42) & (0.30) & (1.08) & (0.46) & (0.33) & (1.10) & (0.44) & (0.31) & (1.11) \\
$\text{SENT}^{LM}$ & 12.2934 & -8.3763 & 23.0555* &  &  &  &  &  &  \\
 & (8.18) & (5.89) & (12.37) &  &  &  &  &  &  \\
$\text{SENT}^A$ &  &  &  & 0.0102 & -0.0086 & -0.0182* &  &  &  \\
 &  &  &  & (0.01) & (0.01) & (0.01) &  &  &  \\
$\hat{\hat{\text{EBP}}}$ &  &  &  &  &  &  & -2.9848** & 1.9784** & -5.6285*** \\
 &  &  &  &  &  &  & (1.24) & (0.89) & (2.12) \\
$\hat{\hat{\text{EBP}}}$-RES &  &  &  &  &  &  & -0.3499 & 0.2840 & -2.6790*** \\
 &  &  &  &  &  &  & (0.42) & (0.30) & (1.01) \\
$\hat{\text{EBP}}$ & -2.4934*** & 1.6449*** & -6.4506*** & -2.1593*** & 1.3231*** & -7.8296*** & -2.1363*** & 1.3648*** & -3.8977*** \\
 & (0.59) & (0.45) & (1.53) & (0.34) & (0.26) & (1.69) & (0.58) & (0.42) & (1.13) \\
$\hat{\text{EBP}}$-RES & -1.4984*** & 1.1422*** & -6.9130*** & -1.7006*** & 1.2750*** & -7.2218*** & -1.1983** & 0.8975** & -4.4098*** \\
 & (0.30) & (0.23) & (1.54) & (0.36) & (0.27) & (1.57) & (0.48) & (0.36) & (1.11) \\
$R^2$ & 0.173 & 0.154 & 0.275 & 0.164 & 0.147 & 0.264 & 0.172 & 0.152 & 0.265 \\
$T$ & 427 & 427 & 427 & 427 & 427 & 427 & 427 & 427 & 427 \\
SHAP(GZF) & 13.33 & 7.27 & 1.33 & 11.71 & 6.50 & 3.16 & 12.95 & 6.98 & 1.48 \\
SHAP($\text{SENT}^{LM}$) & 14.21 & 11.75 & 8.26 &  &  &  &  &  &  \\
SHAP($\text{SENT}^A$) &  &  &  & 9.86 & 11.67 & 5.27 &  &  &  \\
SHAP($\hat{\hat{\text{EBP}}}$) &  &  &  &  &  &  & 15.00 & 12.14 & 9.01 \\
SHAP($\hat{\hat{\text{EBP}}}$-RES) &  &  &  &  &  &  & 4.85 & 4.81 & 11.85 \\
SHAP($\hat{\text{EBP}}$) & 32.43 & 25.96 & 26.00 & 27.30 & 23.50 & 29.70 & 28.46 & 22.20 & 16.55 \\
SHAP($\hat{\text{EBP}}$-RES) & 17.30 & 16.00 & 24.73 & 19.09 & 20.10 & 24.31 & 14.17 & 12.96 & 16.62 \\
\hline
\end{tabular}

%% file: tables/table7-ver2.tex
\begin{tabular}{lcccccc}
\hline\hline
 & \multicolumn{3}{c}{Forecast horizon: 3 months} & \multicolumn{3}{c}{Forecast horizon: 12 months} \\
 & CPI & UER & IPM & CPI & UER & IPM \\
 & (1) & (2) & (3) & (4) & (5) & (6) \\
 \cmidrule(lr){2-4} \cmidrule(lr){5-7}
$\hat{\text{EBP}}$ & -2.0096** & 2.6832*** & -2.5194 & -3.1189*** & 2.3572*** & -2.1268 \\
 & (0.79) & (0.87) & (3.39) & (0.92) & (0.81) & (3.25) \\
BAA-Yield & -0.7455*** & 0.4233* & -0.0393 & -0.8681** & 0.1418 & 0.9453 \\
 & (0.28) & (0.25) & (1.33) & (0.35) & (0.28) & (1.15) \\
TBill-Yield & 0.3858* & 0.0050 & -0.7850 & 0.3619 & 0.2438 & -1.7361* \\
 & (0.21) & (0.20) & (1.07) & (0.26) & (0.22) & (0.98) \\
$R^2$ & 0.354 & 0.184 & 0.084 & 0.297 & 0.187 & 0.067 \\
$T$ & 648 & 521 & 645 & 648 & 521 & 645 \\
SHAP($\hat{\text{EBP}}$) & 11.74 & 32.74 & 9.45 & 21.43 & 37.09 & 8.68 \\
SHAP(BAA-Yield) & 20.93 & 24.15 & 0.74 & 28.66 & 10.43 & 19.46 \\
SHAP(TBill-Yield) & 12.26 & 0.34 & 15.99 & 13.52 & 21.36 & 38.51 \\
\hline
\end{tabular}

%% file: tables/table7b-ver2.tex
\begin{tabular}{lcccc}
\hline\hline
 & \multicolumn{2}{c}{Forecast horizon: 1 quarter} & \multicolumn{2}{c}{Forecast horizon: 4 quarters} \\
 & GDP & NFI & GDP & NFI \\
 & (1) & (2) & (3) & (4)\\
 \cmidrule(lr){2-3} \cmidrule(lr){4-5}
$\hat{\text{EBP}}$ & -4.0475*** & -13.6524*** & -2.9556*** & -8.3668*** \\
 & (1.19) & (2.75) & (0.89) & (2.07) \\
BAA-Yield & -0.0303 & -1.4293** & 0.4170 & 0.8831 \\
 & (0.36) & (0.66) & (0.35) & (0.65) \\
TBill-Yield & -0.6523* & -0.0598 & -0.9605*** & -1.8335*** \\
 & (0.36) & (0.71) & (0.33) & (0.69) \\
$R^2$ & 0.229 & 0.370 & 0.174 & 0.223 \\
$T$ & 104 & 104 & 104 & 104 \\
SHAP($\hat{\text{EBP}}$) & 34.15 & 27.40 & 24.95 & 23.15 \\
SHAP(BAA-Yield) & 1.24 & 13.86 & 17.02 & 11.81 \\
SHAP(TBill-Yield) & 27.99 & 0.61 & 41.24 & 25.80 \\
\hline
\end{tabular}

%% file: tables/table8.tex
\begin{tabular}{lc}
\toprule
 & 12 Month Before \\
\cmidrule(lr){2-2}
November 1890 Panic & 0.635 \\
May 1893 Panic & 0.101 \\
October 1907 Panic & 0.132 \\
November 1930 Panic & 0.209 \\
May 1984 Panic & 0.076 \\
September 2008 Panic & 0.603 \\
NBER Recessions & 0.149 \\
All Other Months & 0.040 \\
\bottomrule
\end{tabular}

%% file: tables/table9.tex
\begin{tabular}{lccccc}
\hline\hline
 & EMP (SA) & UER & IPM & GDP & Recession \\
 & (1) & (2) & (3) & (4) & (5) \\
\cmidrule(lr){2-4} \cmidrule(lr){5-5} \cmidrule(lr){6-6}
TS & 0.1122 & -0.0136 & 0.8356 & 0.0452 & 0.027 \\
 & (0.53) & (0.41) & (0.74) & (0.43) & (0.243) \\
RFF & -0.0139 & 0.1424 & 0.4433 & 0.3021 & 0.146 \\
 & (0.18) & (0.13) & (0.29) & (0.20) & (0.119) \\
GZF-IG & -2.4869*** & 0.9989* & -0.6451 & 0.9463 & 0.914 \\
 & (0.80) & (0.59) & (1.59) & (0.66) & (0.619) \\
$\hat{\text{EBP-IG}}$ & -3.0741*** & 2.4371*** & -10.3530*** & -5.0571*** & 2.334*** \\
 & (0.56) & (0.44) & (2.13) & (0.98) & (0.649) \\
$\hat{\text{EBP-IG}}$-RES & -2.4514*** & 2.0882*** & -11.1571*** & -4.6186*** & 1.405** \\
 & (0.54) & (0.45) & (2.14) & (0.76) & (0.707) \\
$R^2$ & 0.152 & 0.147 & 0.285 & 0.255 & 0.378 \\
$T$ & 408 & 408 & 408 & 136 & 408 \\
SHAP(GZF-IG) & 24.86 & 12.01 & 1.95 & 6.19 & 15.25 \\
SHAP($\hat{\text{EBP-IG}}$) & 30.12 & 28.72 & 30.74 & 33.89 & 42.60 \\
SHAP($\hat{\text{EBP-IG}}$-RES) & 22.81 & 23.37 & 31.47 & 31.27 & 21.74 \\
\hline
\end{tabular}

%% file: tables/table9b.tex
\begin{tabular}{lccccc}
\hline\hline
 & EMP (SA) & UER & IPM & GDP & Recession \\
 & (1) & (2) & (3) & (4) & (5) \\
\cmidrule(lr){2-4} \cmidrule(lr){5-5} \cmidrule(lr){6-6}
TS & 0.1981 & -0.1098 & 1.4008* & 0.2495 & -0.051 \\
 & (0.53) & (0.41) & (0.76) & (0.43) & (0.204) \\
RFF & 0.1113 & 0.0710 & 0.5806 & 0.3503 & 0.083 \\
 & (0.28) & (0.21) & (0.41) & (0.29) & (0.106) \\
GZF-HY & -0.2353 & 0.0010 & 0.3989 & 0.3574 & 0.040 \\
 & (0.36) & (0.26) & (0.69) & (0.42) & (0.277) \\
$\hat{\text{EBP-HY}}$ & -1.8888*** & 1.3898*** & -4.1890*** & -2.6786*** & 1.637*** \\
 & (0.53) & (0.41) & (1.54) & (0.81) & (0.301) \\
$\hat{\text{EBP-HY}}$-RES & -1.2645*** & 0.8384** & -3.3979*** & -1.3984*** & 0.594*** \\
 & (0.48) & (0.34) & (0.87) & (0.42) & (0.198) \\
$R^2$ & 0.166 & 0.146 & 0.221 & 0.193 & 0.393 \\
$T$ & 408 & 408 & 408 & 136 & 408 \\
SHAP(GZF-HY) & 7.40 & 0.05 & 4.08 & 7.69 & 2.32 \\
SHAP($\hat{\text{EBP-HY}}$) & 19.76 & 21.36 & 14.26 & 20.04 & 39.63 \\
SHAP($\hat{\text{EBP-HY}}$-RES) & 35.80 & 34.86 & 31.29 & 27.56 & 40.78 \\
\hline
\end{tabular}

%% file: tables/table3_EBP_base_bkmx.tex
\begin{tabular}{llcc}
\toprule
 &  & Explained Variance & Raw Weight \\
\cmidrule(lr){2-2} \cmidrule(lr){3-4}
\multirow[t]{6}{*}{Top Positive} & Announcements & 14.54 & 9.9099 \\
 & Financial Crisis & 7.92 & 69.6270 \\
 & Financial Markets & 7.29 & 22.0685 \\
 & Financial Intermediaries & 6.20 & 91.8247 \\
 & Corporate Earnings & 1.58 & 4.6189 \\
 & Oil / Mining & 0.86 & 2.6492 \\
\multirow[t]{6}{*}{Top Negative} & National Policies & 22.00 & -22.9425 \\
 & Science / Arts & 18.50 & -59.1553 \\
 & Economic Growth & 12.59 & -31.5703 \\
 & International Relations & 3.39 & -2.9172 \\
 & Technology & 3.38 & -19.2584 \\
 & Industry & 1.74 & -7.4120 \\
\bottomrule
\end{tabular}

%% file: tables/table3_GZF_base_bkmx.tex
\begin{tabular}{llcc}
\toprule
 &  & Explained Variance & Raw Weight \\
\cmidrule(lr){2-2} \cmidrule(lr){3-4}
\multirow[t]{8}{*}{Top Positive} & Announcements & 21.40 & 4.6386 \\
 & Corporate Earnings & 19.44 & 17.1276 \\
 & Financial Crisis & 3.58 & 47.2805 \\
 & Financial Intermediaries & 2.22 & 21.1171 \\
 & Economic Growth & 2.04 & 29.6113 \\
 & International Relations & 1.41 & 11.7146 \\
 & Technology & 0.64 & 4.4471 \\
 & Financial Markets & 0.55 & 3.5570 \\
\multirow[t]{4}{*}{Top Negative} & National Policies & 26.60 & -8.9005 \\
 & Science / Arts & 11.38 & -33.9109 \\
 & Oil / Mining & 8.13 & -13.9849 \\
 & Industry & 2.62 & -4.8444 \\
\bottomrule
\end{tabular}

%% file: tables/table4-ver2-bkmx.tex
\begin{tabular}{lccccc}
\hline\hline
 & EMP (SA) & UER & IPM & GDP & Recession \\
 & (1) & (2) & (3) & (4) & (5) \\
\cmidrule(lr){2-4} \cmidrule(lr){5-5} \cmidrule(lr){6-6}
TS & 0.0584 & -0.1035 & 1.4719* & 0.1511 & -0.098 \\
 & (0.54) & (0.41) & (0.80) & (0.55) & (0.219) \\
RFF & -0.1231 & 0.1904 & 0.3027 & -0.0947 & 0.166 \\
 & (0.30) & (0.23) & (0.48) & (0.37) & (0.179) \\
GZF & -1.5936 & 0.8367 & -1.8229 & 0.1777 & 0.309 \\
 & (0.99) & (0.76) & (1.89) & (0.65) & (0.491) \\
$\hat{\text{EBP}}$-ANNO & -3.6050* & 2.5183* & -9.0015*** & -0.1958 & 2.339** \\
 & (1.86) & (1.40) & (2.70) & (1.32) & (0.971) \\
$\hat{\text{EBP}}$-EARN & -3.5207 & 2.9544 & -4.6317 & -2.2234 & 6.135* \\
 & (5.09) & (3.93) & (10.27) & (4.40) & (3.289) \\
$\hat{\text{EBP}}$-GROWTH & -6.6730** & 4.1798* & -6.9296 & -3.6140* & 3.596*** \\
 & (3.12) & (2.35) & (5.30) & (2.03) & (1.393) \\
$\hat{\text{EBP}}$-FCRIS & -2.3293* & 1.8174 & -1.1548 & -1.1660 & 0.818 \\
 & (1.39) & (1.16) & (4.15) & (2.93) & (0.991) \\
$\hat{\text{EBP}}$-FI & -1.7645* & 1.0890* & -13.4278*** & -4.6653*** & 1.782** \\
 & (0.91) & (0.66) & (3.10) & (1.72) & (0.843) \\
$\hat{\text{EBP}}$-MKT & 1.4108 & 3.0399 & -19.3956** & -0.8944 & -2.583 \\
 & (3.61) & (3.05) & (9.31) & (4.90) & (2.304) \\
$\hat{\text{EBP}}$-IND & 0.3698 & -0.4028 & -11.8411* & -5.3025 & 1.479 \\
 & (4.71) & (3.47) & (6.96) & (4.88) & (1.979) \\
$\hat{\text{EBP}}$-ITNREL & -0.1926 & 0.5868 & -3.2776 & 3.3467 & 0.528 \\
 & (2.02) & (1.21) & (3.44) & (2.50) & (1.059) \\
$\hat{\text{EBP}}$-NTLPLC & -1.4187* & 1.1012** & -6.0239*** & -4.0879*** & 1.414*** \\
 & (0.73) & (0.52) & (1.72) & (0.90) & (0.492) \\
$\hat{\text{EBP}}$-OILMINE & 2.5030 & -1.2673 & -23.6597*** & 7.0906 & 3.608 \\
 & (5.07) & (3.16) & (9.01) & (6.50) & (2.503) \\
$\hat{\text{EBP}}$-SCIART & -2.4208** & 2.1831** & -2.1362 & -2.0079* & 0.810 \\
 & (1.15) & (0.85) & (1.88) & (1.05) & (0.544) \\
$\hat{\text{EBP}}$-TECH & -0.5552 & -0.0339 & -4.7664* & -0.0444 & 1.018 \\
 & (1.38) & (1.14) & (2.45) & (1.16) & (0.782) \\
$\hat{\text{EBP}}$-RES & -1.2318*** & 1.0092*** & -7.2429*** & -2.6480*** & 0.634* \\
 & (0.29) & (0.22) & (1.42) & (0.48) & (0.329) \\
$R^2$ & 0.142 & 0.130 & 0.305 & 0.167 & 0.394 \\
$T$ & 427 & 427 & 427 & 142 & 427 \\
\hline
\end{tabular}

%% file: appendix.tex
\clearpage
\newpage
\appendix
{\centering
\section*{Understanding the Excess Bond Premium \\ \vspace{.25in} Appendix}}
\vspace{.35in}

\section{Technical Details}\label{app:technical}

\noindent BKMX refers to \citet{bkmx:21} and LM refers to \citet{LM:11}.

\subsection{Data series}
We obtain the following series from the St. Louis Fed FRED system:
\begin{itemize}
    \item Slope of the Treasury yield curve, 10-year minus 3-month (FRED:T10Y3MM)
    \item Real federal funds rate: the average effective federal funds rate in month $t$ (FRED: FEDFUND) minus realized inflation, where realized inflation is the log-change in the core PCE price index between month $t-1$ and its lagged value one year earlier (FRED: PCEPILFE)
    \item Private nonfarm payroll employment (FRED: PAYNSA, PAYEMS)
    \item Unemployment (FRED: UNRATE)
    \item Industrial production (FRED: IPMAN)
\end{itemize}

\subsection{Text Pre-processing and Vectorization}
To retrieve articles from TDM Studio, we use the following steps:
\begin{enumerate}
    \item For 1889-07-08 to 2013-12-31, we use the two datasets: ``Wall Street Journal (1889-1922)'' (ID: 55361) and ``Wall Street Journal (1923-2013)'' (ID: 45441). Set the document type to be Front Page/Cover Story. This gives a total of 247,607 articles.
    \item For 2014-01-01 to 2023-12-31, we use the dataset ``Wall Street Journal (1984-9999)'' (ID: 10482). Retrieve all articles. Screen the articles' GOIDs by ``Start Page==A.01'' or ``Start Page==A.1''. This gives a total of 22,368 articles. 
    \item Then we retrieve from TDM Studio the news text. Remove the articles that do not contain any text. This gives 269,903 articles in the end.
    \item \textbf{Note}: the news data from 1892-01-01 to 1892-06-01 are missing. This is potentially a misclassification from TDM. 
\end{enumerate}

\noindent We apply the following pre-processing steps:
\begin{enumerate}
    \item Replace newlines and any whitespace characters with a single space.
    \item Unescape HTML characters. Remove URLs and email addresses.
    \item Remove all numerical values, \$ . / \% symbols, multiple spaces, and leading/tailing spaces.
    \item Tokenize the text, remove digits, stop words and puntuations.
    \item Lemmatize the text.
    \item More filtering on bigrams and trigrams: [``year year'', ``month month'', ``week week'', ``day day'', ``wall street journal'', ``new york times'', ``new york'', ``dow jones newswires''].
    \item Cleanup any whitespace again.
\end{enumerate}

\noindent We vectorize each individual article based on the BKMX dictionary, after which we reduce to 180 topics. For sentiment, we use the LM dictionary and then reduce to positive/negative sentiment counts.

\subsection{Construction of $\hat{EBP}$ and related variables}

\subsubsection{$\hat{EBP}$}
Specifically, for a given month $t$ and topic $k$, the estimated attention that news articles in month $t$ allocate to topic $k$ is determined by the frequency of terms associated with $k$:
\begin{align*}
    \hat{\theta}_{k,t} = \frac{\sum_{i=1}^{N_t}\mathbbm{1}(\hat{z}_{i,t}=k)}{\sum_{q=1}^K\sum_{i=1}^{N_t}\mathbbm{1}(\hat{z}_{i,t}=q)},
\end{align*}
where $\hat{z}_{i,t}$ represents topic assignment of word $i$ in the total vocabulary for month $t$, $K=180$ represents the total number of topics, and $N_t$ is the total vocabulary count in all articles in month $t$. Note that we pool together all articles in a month.

Using $\hat{\theta}_{k,t}$ and $s_t$, we represent the monthly articles as a vectorized feature matrix $X\in\mathbb{R}^{T\times N}$, where $T$ is the time-series length, $N$ is number of features. For $\hat{\theta}_{k,t}$, $N=180$, and for $s_t$, $N=2$. We then apply lasso regression to predict $y=$EBP:
\begin{equation}\label{eq:lasso}
    \arg\min_w \frac{1}{2T} \|Xw+\alpha-y\|_2^2 + \lambda \|w\|_1,
\end{equation} 
where $w\in\mathbb{R}^{N}$ is the weight vector for each feature in $X$, and $\alpha$ is the constant intercept. 

To ensure the robustness of the model, we employ five-fold cross-validation and introduce an additional regularization factor of $\lambda=10^{-5}$ to the L1 loss.

We start by training on an initial window of 180 months (from 1973-01 to 1987-12). We then employ an expanding window approach that progressively increases the data used for future predictions. In the ``in-sample'' approach, we use all data through month $t$ to train the model and then use the text from month $t$ to predict $EBP_t$. In the ``out-of-sample'' approach, we train the model using data through month $t-1$ and then use the text from month $t$ to predict $EBP_t$. This latter approach effectively uses a stale model at each step.

As noted in \cite{nvix}, due to the high dimensionality of features (180), the model’s weights may be less reliable with shorter training windows (such as 180 months). Thus, early in the process, the discrepancy between in-sample and out-of-sample data can be significant. However, as the training period lengthens, this discrepancy diminishes, and the weight estimates become more stable.

We employ this same approach when constructing $\hat{GZF}$, the comp
This same expanding window approach is used in constructing $\hat{\text{GZF}}$, where $GZF$ equals the non-decomposed average credit spread minus EBP.

\subsubsection{Sentiment-Based Variables}

To calculate $SENT^{LM}$, we calculate the polar sentiment using positive and negative word counts in each article $a$. Specifically, we count the positive sentiment words $c_a^+$ and negative sentiment words $c_a^-$ in each article, with sentiment determined by the LM dictionary\footnote{Online at \url{https://sraf.nd.edu/loughranmcdonald-master-dictionary}.}. The sentiment score of an article $a$ equals:
\begin{align*}
    s_a &= \frac{c_a^+ - c_a^-}{c_a^+ + c_a^-}.
\end{align*}
For any month $t$, we compute the monthly sentiment $s_t$ by aggregating $c_a^+$ and $c_a^-$ across all articles in that month.

We construct a new sentiment time series, $\text{SENT}^A$, that equals the weighted average of sentiment across different topics, where the weights themselves are the attention weights from our estimation of $\hat{EBP}$. To calculate the sentiment associated with each topic $k$ in a given month $t$, which we denote $s_{k,t}$, we first calculate the article-level sentiment, multiply it against the attention share of that article to each topic, and then add up over that product over articles. Specifically, we start by computing a monthly sentiment score for each topic, $s_{k,t}$, as follows (where we omit the time subscripts):
\begin{enumerate}
    \item For each article $a$ in the set of articles $A_t$ in month $t$, calculate sentiment $s_a$;
    \item For each article $a$, determine the BKMX topic attention $\hat{\theta}_{a,k}$ for each topic $k$, such that $\sum_k \hat{\theta}_{a,k} = 1$;
    \item Calculate the monthly topic attention $s_{k,t}$ for topic $k$ in month $t$ as $s_{k,t} = \sum_{a\in A_t} s_a \hat{\theta}_{a,k}$.
\end{enumerate}

The sentiment time series equals the weighted average of $s_{k,t}$ across topics, where the weights equal the attention weights $w_{k,t-1}$ from the estimation of $\hat{EBP}$:
\begin{align*}
    \text{SENT}_t^A &= \sum_{k\in M} w_{k,t-1} s_{k,t},
\end{align*}

We construct $\hat{\hat{\text{EBP}}}$ following our identical process to $\hat{EBP}$ except where we consider $SENT^{LM}$ as the lone ``topic'' that the model can apply a weight to each month. Effectively, $\hat{\hat{\text{EBP}}}$ is a rolling-window projection of EBP onto $SENT^{LM}$.

\subsection{SHAP Decomposition}\label{subsec:SHAP}
In order to assess the contribution of any given variable to our prediction model, we conduct a Shapley decomposition of the predicted value. The basic intuition behind the Shapley decomposition is that it provides the average contribution of a given variable to the predicted value, averaged across all permutations of predictor variables.

With the game-theoretical foundation established by Shapley (\cite{Shapley1951, Shapley1988AVF}), \cite{shapexact} propose the use of Shapley regression values to explain model predictions.
In this method, to compute the feature importance of $x_i$, the model is trained on all possible feature subsets $S\subseteq \{x_1,...,x_n\}\setminus \{x_i\}$. Let $y_{S\cup \{x_i\}}$ represent the outcome of the model when trained with the feature $x_i$, and $y_{S}$ denote the outcome when the model is trained without the feature $x_i$. 
The difference in the outcomes, $y_{S\cup \{x_i\}} - y_{S}$, is computed for all possible subsets $S\subseteq \{x_1,...,x_n\}\setminus \{x_i\}$. The Shapley value, computed as the weighted average of all possible differences, is then used to determine the feature importance of $x_i$:
\begin{equation}
\phi_i = \sum_{S\subseteq \{x_1,...,x_n\}\setminus \{x_i\}} \frac{|S|!(n-1-|S|)!}{n!} [y_{S\cup \{x_i\}} - y_{S}]
\end{equation}

SHAP (SHapley Additive exPlanations), developed by \cite{SHAP}, is a comprehensive framework designed for interpreting predictions. 
Within this framework, the Shapley value explanation is depicted as an additive feature attribution method, which can be represented by the following linear explanation model:
\begin{equation}
g(z')=\phi_0 + \sum_{i=1}^M \phi_i z_i',
\end{equation}
where $g$ denotes the explanation model, and $z'\in \{0, 1\}^M$ represents the simplified features, also known as the coalition vector. In this context, $z_i'=1$ if feature $i$ is present, and $z_i'=0$ if the feature is absent. $M$ is the maximum coalition size, while $\phi_i$ is the Shapley value for a feature $i$. 
Given $x=(x_1,...,x_n)$ as the full feature vector, the coalition vector $x'=\mathbf{1}$ as a vector of all ones, the explanation can be expressed as: 
\begin{equation}
g(x')=\phi_0 + \sum_{i=1}^M \phi_i,
\end{equation}
which is the sum of Shapley values for all features $x_i$. 

In this framework, features with large absolute Shapley values are deemed important. We calculate the average of the absolute Shapley value of that feature across the dataset:
\begin{equation}
I_j = \frac{1}{N} \sum_{i=1}^{N} |\phi_j^{(i)}|, 
\end{equation}
where $N$ represents the size of the dataset, and $j$ is the feature for which we wish to calculate the importance. Subsequently, the SHAP value is normalized as follows:
\begin{equation}
    \hat{I}_j = \frac{I_j}{\sum_{k=1}^n I_k} \times 100
\end{equation}

\subsection{Metatopic Classification}\label{sec:gptmeta}
To categorize the 180 BKMX topics into broader metatopics for streamlined analysis, our main analysis employs GPT-o1. We prompt GPT-o1 to categorize our topics into ten meta-topics. We ask GPT-o1 to explicitly incorporate sentiment as well. Our prompt is as follows:
\small
\begin{verbatim}
I want to condense these topics into 10 broader metatopics. 
Also, split the positive/negative topics. 
For example, "Optimism", "Record High", "Positive Sentiment" are sentiment 
related and positive;
"major concern" is sentiment related, 
but negative; "Options/VIX", "Treasury bonds", "NASD are financial market words, 
and "Bush / Obama / Trump", "Reagan" are politics related, 
"Disease", "Public health" should be health relavent. 
Create a json dictionary of the form:
{
"topic1": ["word1", "word2"],
"topic2": ["word3", "word3"],
}
Note that each of the topics should be in one and only one higher metatopic. 
Make sure that all of these topics are included. 
You can create a "Other" category for words that doesn't fit into any of 
the categories you have. 
A potential list of metatopics: ["Positive Sentiment", "Negative Sentiment", 
"Financial Intermediaries", "Financial Markets / Economy", 
"Industry", "Economic Growth", 
"Science / Technology",  "Crisis / Disasters", 
"Politics", "Other"]

----
<180 BKMX topics>
\end{verbatim}
\normalsize
After receiving the response from GPT-o1, we manually added any missing topics and assigned the topic ``Financial crisis'' to its own separate category.

We calculate the aggregate weight of a metatopic $M$ as the sum of the weights of all topics within that metatopic:
\begin{align*}
    w_M = \sum_{k\in M} w_{k}
\end{align*}
where $w_k$ represents the raw weight of topic $k$ as determined by the Lasso regression.

To quantify the contribution of each metatopic, we compute its explained variance:
\begin{align*}
    h(M) &= \frac{\sum_{i\in M}\sum_{j\in M} w_i w_j \text{Cov}(f_i, f_j)}{\sum_M\sum_{i\in M}\sum_{j\in M} w_i w_j \text{Cov}(f_i, f_j)} \times 100,
\end{align*}
where $f_i$ denotes the topic attention, represented by $\hat{\theta}_{i}$. The explained variance is normalized as a percentage.

We also define $\hat{\text{EBP}}$-M for a specific metatopic $M$, as well as the residuals, using the following equations:
\begin{align*}
    \hat{\text{EBP}}\text{-M} &= \sum_{k\in M} w_k f_k\\
    \hat{\text{EBP}}\text{-M-RES} &= \text{EBP} - \hat{\text{EBP}}\text{-M}\\
    \hat{\text{EBP}}\text{-RES} &= \text{EBP} - \hat{\text{EBP}}
\end{align*}

In our robustness analysis, we explicitly follow the meta-topic categorization of \citet{bkmx:21} Figure 2. We employ a high-level meta-topic categorization of 12 topics instead of the 23 listed in their Internet Appendix for parsimony.

\clearpage
\newpage
\renewcommand{\thetable}{B.\arabic{table}}
\setcounter{table}{0}
\renewcommand{\thefigure}{B.\arabic{figure}}
\setcounter{figure}{0}
\section{Additional Tables \& Figures}
\begin{table}[!htb]
\caption{Predicting Macroeconomic Fluctuations (12 Month/4 Quarter)}\label{tab:prediction-macro-fluctuations-12m}
{
This table reports the regression coefficients and errors of the specification reported in Table~\ref{tab:prediction-macro-fluctuations}, but with a 12-month/4-quarter forecast horizon.
}

\bigskip
\centering

A. $\hat{\text{EBP}}$\\
\resizebox*{0.8\textwidth}{!}{\input{tables/table2-12m-ver2.tex}}

\vspace{.3cm}
B. EBP\\
\resizebox*{0.8\textwidth}{!}{\input{tables/table2b-12m.tex}}

\end{table}

\clearpage
\newpage
\begin{table}[!htb]
\caption{Metatopic Classification (GPT)}\label{tab:metatopic-classification}

\bigskip
\centering
\resizebox*{1.0\textwidth}{!}{\input{tables/metatopics.tex}}

\end{table}

\clearpage
\newpage
\begin{table}[!htb]
\caption{Metatopic Classification (BKMX)}\label{tab:metatopic-classification-bkmx}

\bigskip
\centering
\resizebox*{1.0\textwidth}{!}{\input{tables/metatopics_bkmx.tex}}

\end{table}

\clearpage
\newpage
\begin{table}[!htb]
\caption{Historical Regression Topic Decomposition (GPT Monthly)}\label{tab:historical-topic-decomp-gpt}
{\small 
This table reports the regression coefficients and standard errors of the three- and twelve-month-ahead macroeconomic forecasts in Table \ref{tab:backward-projection} Panel A, replacing $\hat{EBP}$ with the excess bond premiums driven by each of the 11 metatopics categorized by GPT-o1.
}
\bigskip

{\centering
\resizebox*{1.0\textwidth}{!}{\input{tables/table7_gpt_decomp}}
}
\end{table}

\clearpage
\newpage
\begin{table}[!htb]
\caption{Historical Regression Topic Decomposition (GPT Quarterly)}\label{tab:historical-topic-decomp-gptq}
{\small 
This table reports the regression coefficients and standard errors of the one- and four-quarter-ahead macroeconomic forecasts in Table \ref{tab:backward-projection} Panel B, replacing $\hat{EBP}$ with the excess bond premiums driven by each of the 11 metatopics categorized by GPT-o1.
}

\bigskip
\centering
\resizebox*{1.0\textwidth}{!}{\input{tables/table7b_gpt_decomp}}

\end{table}

%% file: tables/table2-12m-ver2.tex
\begin{tabular}{lccccc}
\hline\hline
 & EMP (SA) & UER & IPM & GDP & Recession \\
 & (1) & (2) & (3) & (4) & (5) \\
\cmidrule(lr){2-4} \cmidrule(lr){5-5} \cmidrule(lr){6-6}
TS & 0.3451 & -0.2021 & 1.2608*** & 0.1465 & -0.644*** \\
 & (0.36) & (0.22) & (0.39) & (0.36) & (0.227) \\
RFF & -0.0271 & 0.1188 & 0.3440 & 0.0158 & 0.127 \\
 & (0.17) & (0.11) & (0.27) & (0.20) & (0.135) \\
GZF & -0.7511 & 0.2996 & -0.4506 & -0.7626 & 0.553 \\
 & (0.51) & (0.31) & (1.24) & (0.77) & (0.666) \\
$\hat{\text{EBP}}$ & -1.8560*** & 1.0193*** & -2.9791*** & -1.3925*** & 1.804*** \\
 & (0.31) & (0.17) & (1.00) & (0.37) & (0.441) \\
$\hat{\text{EBP}}$-RES & -1.4050*** & 0.9644*** & -4.1379*** & -1.6651*** & 1.119*** \\
 & (0.27) & (0.19) & (0.79) & (0.49) & (0.356) \\
$R^2$ & 0.272 & 0.276 & 0.240 & 0.165 & 0.426 \\
$T$ & 427 & 427 & 427 & 142 & 427 \\
SHAP(GZF) & 15.66 & 8.59 & 3.64 & 18.53 & 9.53 \\
SHAP($\hat{\text{EBP}}$) & 35.41 & 26.73 & 22.04 & 32.60 & 33.42 \\
SHAP($\hat{\text{EBP}}$-RES) & 23.79 & 22.44 & 27.17 & 34.12 & 17.31 \\
\hline
\end{tabular}

%% file: tables/table2b-12m.tex
\begin{tabular}{lccccc}
\hline\hline
 & EMP (SA) & UER & IPM & GDP & Recession \\
 & (1) & (2) & (3) & (4) & (5) \\
\cmidrule(lr){2-4} \cmidrule(lr){5-5} \cmidrule(lr){6-6}
TS & 0.2776 & -0.1938 & 1.4453*** & 0.1794 & -0.537** \\
 & (0.34) & (0.21) & (0.39) & (0.35) & (0.230) \\
RFF & -0.0521 & 0.1217 & 0.4143 & 0.0259 & 0.155 \\
 & (0.17) & (0.10) & (0.25) & (0.20) & (0.130) \\
GZF & -0.8988* & 0.3178 & -0.0552 & -0.6785 & 0.716 \\
 & (0.46) & (0.27) & (1.16) & (0.74) & (0.598) \\
EBP & -1.6162*** & 0.9904*** & -3.6378*** & -1.5407*** & 1.430*** \\
 & (0.24) & (0.15) & (0.72) & (0.38) & (0.344) \\
$R^2$ & 0.269 & 0.277 & 0.234 & 0.170 & 0.409 \\
$T$ & 427 & 427 & 427 & 142 & 427 \\
SHAP(GZF) & 25.35 & 11.80 & 0.54 & 22.93 & 14.90 \\
SHAP(EBP) & 41.84 & 33.73 & 32.87 & 51.25 & 39.55 \\
\hline
\end{tabular}

%% file: tables/metatopics.tex
\begin{tabular}{ll}
\toprule
 & subtopics \\
 \cmidrule(lr){2-2}
 Positive Sentiment & Record high, Positive sentiment, Optimism, Agreement reached, Revenue growth \\
 Negative Sentiment & Problems, Job cuts, Earnings losses, Bankruptcy, Major concerns,\\
 & Indictments, Challenges, Corrections / amplifications \\
 Financial Intermediaries & M\&A, Savings \& loans, IPOs, Nonperforming loans, Credit ratings,\\
 & Control stakes, Mutual funds, Venture capital, Drexel, Investment banking,\\
 & International exchanges, Bank loans, Mortgages, Acquired investment banks, Credit cards, \\
 & Insurance, Private equity / hedge funds, Buffett, Pensions, Accounting \\
 Financial Markets / Economy & Profits, Bond yields, Short sales, Federal Reserve, Small caps, \\
 & Treasury bonds, SEC, Futures / indices, Exchanges / composites, Currencies / metals,\\
 & Financial reports, Bear / bull market, Earnings forecasts, Oil market, Commodities,\\
 & Convertible / preferred, Macroeconomic data, Options / VIX, Trading activity, NASD, \\
 & Earnings, Fees, Product prices, Rental properties, Subsidiaries, \\
 & Share payouts, Management changes, Corporate governance, Executive pay, Takeovers, \\
 & European sovereign debt, Revised estimate \\
 Industry & Soft drinks, Electronics, Steel, Cable, Fast food,\\
 & Music industry, Broadcasting, Chemicals / paper, Mining, Pharma,\\
 & Publishing, Aerospace / defense, Phone companies, Tobacco, Automotive,\\
 & Movie industry, Machinery, Oil drilling, Rail / trucking / shipping, Airlines,\\
 & Health insurance, Retail, Couriers, Utilities, Foods / consumer goods,\\
 & Luxury / beverages, Casinos, Agriculture, Real estate\\
 Economic Growth & Economic growth, Small business, Competition\\
 Science / Technology & Internet, Mobile devices, Research, Computers, Biology / chemistry / physics,\\
 & Space program, Microchips, Systems, Software\\
 Crisis / Disasters & Natural disasters, Police / crime, Disease, Environment, Recession, Terrorism\\
 Financial Crisis & Financial crisis\\
 Politics & Economic ideology, Middle east, US defense, Political contributions, Justice Department,\\
 & Regulation, Unions, Private / public sector, Russia, Trade agreements,\\
 & Latin America, Japan, Nuclear / North Korea, NY politics, State politics,\\
 & Immigration, US Senate, Government budgets, Courts, Safety administrations,\\
 & Reagan, Bush / Obama / Trump, Taxes, Iraq, National security,\\
 & Elections, European politics, Mexico, UK, Committees,\\
 & Clintons, China, Canada / South Africa, France / Italy, Germany,\\
 & Southeast Asia, Watchdogs, Activists, California, Lawsuits\\
 Other & Changes, Mid-size cities, Scenario analysis, Restraint, Key role,\\
 & News conference, Announce plan, C-suite, Company spokesperson, Programs / initiatives,\\
 & People familiar, Sales call, Cultural life, Marketing, Arts,\\
 & Small changes, Small possibility, Spring / summer, Humor / language, Mid-level executives,\\
 & Negotiations, Size, Long / short term, Wide range, Connecticut,\\
 & Schools, Gender issues\\
\bottomrule
\end{tabular}

%% file: tables/metatopics_bkmx.tex
\begin{tabular}{ll}
\toprule
    & subtopics \\
    \cmidrule(lr){2-2}
Financial Intermdiaries & NASD, Accounting, Acquired investment banks, Investment banking, Private equity / hedge funds, \\
& Mutual funds, Bankruptcy, SEC, Corporate governance, Drexel, \\
& Control stakes, Real estate, M\&A, Convertible / preferred, Takeovers, \\
& Bank loans, Credit ratings, Mortgages, Nonperforming loans, Savings \& loans, Financial crisis \\
Financial Markets & Bear / bull market, Share payouts, IPOs, Short sales, Treasury bonds, Bond yields, \\
& Options / VIX, Exchanges / composites, Commodities, Currencies / metals, Trading activity, \\
& International exchanges, Small caps \\
Economic Growth & European sovereign debt, Federal Reserve, Macroeconomic data, Economic growth, Optimism, \\
& Record high, Recession, Product prices \\
Oil \& Mining & Mining, Steel, Machinery, Agriculture, Oil market, Oil drilling \\
Corporate Earnings & Earnings, Profits, Earnings forecasts, Earnings losses, Financial reports, \\
& Revised estimate, Small changes \\
Industry & Venture capital, Small business, Subsidiaries, Chemicals / paper, Revenue growth, \\
& Luxury / beverages, Soft drinks, Foods / consumer goods, Competition, Casinos, \\
& Fast food, Couriers, Credit cards, Tobacco, Cable, Insurance \\
Technology & Phone companies, Internet, Software, Computers, Microchips, \\
& Electronics, Mobile devices \\
National Policies & Fees, Executive pay, Pensions, Health insurance, Taxes, \\
& Government budgets, Unions, Job cuts, Mid-level executives, Connecticut, \\
& Management changes, C-suite, Retail, Automotive, Aerospace / defense, US defense, \\
& Airlines, Pharma, Disease, Rail / trucking / shipping, Natural disasters, Police / crime, \\
& Rental properties, NY politics, California, Mid-size cities, Environment, Regulation, \\
& Utilities, Private / public sector, Political contributions, State politics, National security, Watchdogs, \\
& Safety administrations, Lawsuits, Courts, Indictments, Justice Department, European politics, \\
& Elections, US Senate, Bush / Obama / Trump, Reagan, Clintons \\
International Relations & UK, Canada / South Africa, France / Italy, Germany, Japan, \\
& Trade agreements, Latin America, Russia, Southeast Asia, China, \\
& Iraq, Nuclear / North Korea, Terrorism, Middle east \\
Science \& Arts & Arts, Cultural life, Gender issues, Humor / language, Positive sentiment, \\
& Sales call, Immigration, Schools, Economic ideology, Publishing, \\
& Broadcasting, Movie industry, Music industry, Marketing, Scenario analysis, \\
& Research, Wide range, Size, Space program, Biology / chemistry / physics, \\
& Systems, Programs / initiatives, Challenges, Key role, Problems, \\
& Spring / summer, Changes, Long / short term, Small possibility \\
Announcements & Committees, Restraint, News conference, Negotiations, Agreement reached, \\
& People familiar, Company spokesperson, Mexico, Activists, Corrections / amplifications, \\
& Buffett, Major concerns, Futures / indices, Announce plan \\
\bottomrule
\end{tabular}
    

%% file: tables/table7_gpt_decomp.tex
\begin{tabular}{lcccccc}
\hline\hline
 & \multicolumn{3}{c}{Forecast horizon: 3 months} & \multicolumn{3}{c}{Forecast horizon: 12 months} \\
 & CPI & UER & IPM & CPI & UER & IPM \\
 & (1) & (2) & (3) & (4) & (5) & (6) \\
 \cmidrule(lr){2-4} \cmidrule(lr){5-7}
BAA-Yield & -0.9107*** & 0.3939 & -0.6091 & -0.9550*** & 0.1191 & 0.6115 \\
 & (0.28) & (0.24) & (1.29) & (0.33) & (0.22) & (1.20) \\
TBill-Yield & 0.5760** & 0.1697 & -1.1041 & 0.5091 & 0.4859** & -1.8014 \\
 & (0.28) & (0.17) & (1.14) & (0.36) & (0.21) & (1.13) \\
$\hat{\text{EBP}}$-CRISDIS & -3.9096* & 5.8426** & -26.3199** & -6.0964** & 1.5286 & -1.4399 \\
 & (2.26) & (2.30) & (11.92) & (2.53) & (1.80) & (9.37) \\
$\hat{\text{EBP}}$-GROWTH & -22.7023 & 13.0280 & -116.0916** & -30.4708 & 27.0302*** & -108.8825** \\
 & (16.67) & (9.34) & (56.59) & (21.42) & (10.35) & (48.90) \\
$\hat{\text{EBP}}$-FCRIS & 15.1853 & 3.2918 & 75.3898 & 16.0644* & -7.2919 & 39.8966* \\
 & (12.31) & (12.17) & (57.91) & (8.25) & (7.47) & (23.59) \\
$\hat{\text{EBP}}$-FI & 2.0039 & -0.5861 & 34.6423* & 6.7004 & -5.5512 & 29.4241** \\
 & (5.72) & (5.32) & (20.22) & (5.19) & (3.52) & (13.93) \\
$\hat{\text{EBP}}$-MKTECO & 0.5785 & 0.6038 & 8.9168 & 2.4912 & 1.4745 & 3.7208 \\
 & (2.06) & (2.70) & (8.81) & (2.20) & (1.57) & (6.88) \\
$\hat{\text{EBP}}$-IND & -3.0246 & 7.6833*** & -8.0325 & -4.5291** & 11.0967*** & -9.1846 \\
 & (2.75) & (2.98) & (12.00) & (2.10) & (3.48) & (8.72) \\
$\hat{\text{EBP}}$-NEGSENT & -8.1665** & 5.2732 & -19.3829 & -9.9961*** & 1.7103 & 5.3793 \\
 & (3.89) & (3.24) & (15.07) & (3.86) & (3.06) & (18.10) \\
$\hat{\text{EBP}}$-OTHER & 2.7964 & 2.8696* & 10.6970 & 1.9130 & 1.6178 & 3.2331 \\
 & (2.31) & (1.59) & (6.58) & (2.25) & (1.23) & (6.02) \\
$\hat{\text{EBP}}$-PLTC & -5.7514*** & 0.8278 & -5.9607 & -6.6927*** & 2.3359** & -7.2438 \\
 & (1.82) & (1.76) & (6.91) & (1.83) & (1.10) & (4.62) \\
$\hat{\text{EBP}}$-POSSENT & 0.4384 & -5.0143 & -3.7939 & 1.0625 & -0.1558 & 1.5047 \\
 & (2.92) & (4.19) & (17.76) & (3.62) & (4.94) & (17.20) \\
$\hat{\text{EBP}}$-SCITECH & 8.1822 & 9.2513 & 7.4003 & -10.2163 & 10.8088 & 2.0245 \\
 & (8.32) & (6.82) & (29.09) & (7.88) & (6.78) & (24.43) \\
$R^2$ & 0.378 & 0.212 & 0.114 & 0.379 & 0.269 & 0.096 \\
$T$ & 648 & 521 & 645 & 648 & 521 & 645 \\
\hline
\end{tabular}

%% file: tables/table7b_gpt_decomp.tex
\begin{tabular}{lcccc}
\hline\hline
 & \multicolumn{2}{c}{Forecast horizon: 1 quarter} & \multicolumn{2}{c}{Forecast horizon: 4 quarters} \\
 & GDP & NFI & GDP & NFI \\
 & (1) & (2) & (3) & (4)\\
 \cmidrule(lr){2-3} \cmidrule(lr){4-5}
BAA-Yield & -0.0638 & -0.0990 & 0.0781 & 0.6247 \\
 & (0.41) & (1.00) & (0.31) & (0.77) \\
TBill-Yield & -0.8668** & -1.7910** & -0.8557*** & -2.2642*** \\
 & (0.41) & (0.88) & (0.26) & (0.85) \\
$\hat{\text{EBP}}$-CRISDIS & -9.9536** & -46.1716*** & -1.4048 & -23.4007*** \\
 & (4.11) & (9.74) & (2.25) & (5.89) \\
$\hat{\text{EBP}}$-GROWTH & -35.4962** & -74.2919* & -24.5612*** & -61.1542** \\
 & (14.56) & (42.68) & (6.83) & (24.24) \\
$\hat{\text{EBP}}$-FCRIS & 12.6895 & 61.2671 & -10.3102 & 62.3121 \\
 & (22.13) & (64.66) & (18.10) & (45.33) \\
$\hat{\text{EBP}}$-FI & -0.4373 & 18.4033 & 6.0038 & 32.9477* \\
 & (7.49) & (21.19) & (5.59) & (17.59) \\
$\hat{\text{EBP}}$-MKTECO & 2.9484 & 18.7287 & -6.3378* & -2.0076 \\
 & (4.59) & (13.47) & (3.23) & (8.58) \\
$\hat{\text{EBP}}$-IND & 3.4128 & -3.8337 & -3.3241 & -19.6214** \\
 & (5.49) & (15.66) & (3.61) & (8.96) \\
$\hat{\text{EBP}}$-NEGSENT & 5.9970 & -18.8292 & 18.0304*** & 13.9432 \\
 & (7.95) & (21.12) & (5.07) & (16.40) \\
$\hat{\text{EBP}}$-OTHER & -4.5136** & -3.8281 & -4.6379** & -2.2768 \\
 & (2.17) & (7.28) & (2.10) & (5.43) \\
$\hat{\text{EBP}}$-PLTC & -5.5104** & -14.1727** & -3.5432** & -9.4743*** \\
 & (2.48) & (6.91) & (1.43) & (3.61) \\
$\hat{\text{EBP}}$-POSSENT & -6.9328 & -7.3232 & 2.0570 & 9.5250 \\
 & (7.13) & (19.10) & (4.28) & (10.66) \\
$\hat{\text{EBP}}$-SCITECH & 8.3259 & 25.4762 & 7.2551 & 3.3264 \\
 & (12.75) & (34.77) & (6.55) & (22.86) \\
$R^2$ & 0.221 & 0.434 & 0.315 & 0.278 \\
$T$ & 104 & 104 & 104 & 104 \\
\hline
\end{tabular}